\documentclass[10pt,letterpaper]{article}
\usepackage[top=0.85in,right=2.75in,footskip=0.75in]{geometry}

\usepackage{amsmath,amssymb}

\usepackage{changepage}

\usepackage[utf8x]{inputenc}

\usepackage{textcomp,marvosym}

\usepackage{cite}
\usepackage[final]{changes}

\usepackage{nameref,hyperref}

\usepackage[left]{lineno}

\usepackage{microtype}
\DisableLigatures[f]{encoding = *, family = * }

\usepackage{array}

\newcolumntype{+}{!{\vrule width 2pt}}

\newlength\savedwidth

\usepackage{setspace} 
\raggedright
\usepackage[aboveskip=1pt,labelfont=bf,labelsep=period,justification=raggedright,singlelinecheck=off]{caption}

\makeatletter
\renewcommand{\@biblabel}[1]{\quad#1.}
\makeatother

\usepackage{lastpage,fancyhdr,graphicx}
\usepackage{epstopdf}
\fancyheadoffset[R]{2.25in}
\usepackage{ifthen}
\newboolean{isdraft}
\setboolean{isdraft}{false}
\ifthenelse{\boolean{isdraft}}{%

  \newcounter{comments}
  \newcommand{\ananthan}[1]{\addtocounter{comments}{1}{\color{orange}[AN \thecomments: #1]}}
  \newcommand{\mb}[1]{\addtocounter{comments}{1}{\color{green}[MB \thecomments: #1]}}
  \newcommand{\tobias}[1]{\addtocounter{comments}{1}{\color{blue}[TB \thecomments: #1]}}
  \newcommand{\new}[1]{{\color{red}#1}}

}{
\newcommand{\ananthan}[1]{}
\newcommand{\mb}[1]{}
\newcommand{\tobias}[1]{}
\newcommand{\new}[1]{#1}
}

\definechangesauthor[name={Ananthan}, color=red]{AN}
\definechangesauthor[name={Tobias}, color=red]{TR}
\definechangesauthor[name={Tobias}, color=red]{MB}
\definechangesauthor[name={John}, color=red]{JD}
\setauthormarkup{}

\begin{document}
\vspace*{0.2in}

\begin{flushleft}
{\Large
\textbf\newline{Dropping diversity of products of large US firms: Models and measures} 
}
\newline
\\
Ananthan Nambiar\textsuperscript{1,2\P},
Tobias Rubel\textsuperscript{3,4\P},
James McCaull\textsuperscript{5},
Jon deVries\textsuperscript{4},
Mark Bedau\textsuperscript{3*}
\\
\bigskip
\textbf{1} Department of Bioengineering, University of Illinois at Urbana-Champaign, Urbana, Illinois, USA
\\
\textbf{2} Carl R. Woese Institute for Genomic Biology, University of Illinois at Urbana-Champaign, Urbana, Illinois, USA
\\
\textbf{3} Department of Philosophy, Reed College, Portland, Oregon, USA
\\
\textbf{4} Department of Biology, Reed College, Portland, Oregon, USA
\\
\textbf{5} Department of Computer Science, Reed College, Portland, Oregon, USA
\bigskip

%
%
    \P These authors contributed equally to this work.

* Corresponding Author

Email: mab@reed.edu (MB)
\pagebreak 

\end{flushleft}
\section*{Abstract}
It is widely assumed that in our lifetimes the products available in the global economy have become more diverse. This assumption is difficult to investigate directly, however, because it is difficult to collect the necessary data about every product in an economy each year. We solve this problem by mining publicly available textual descriptions of the products of every large US firms each year from 1997 to 2017. Although many aspects of economic productivity have been steadily rising during this period, our text-based measurements show that the diversity of the products of at least large US firms has steadily declined. This downward trend is visible using a variety of product diversity metrics, including some that depend on a measurement of the similarity of the products of every single pair of firms. The current state of the art in comprehensive and detailed firm-similarity measurements is a Boolean word vector model due to Hoberg and Phillips. We measure diversity using firm-similarities from this Boolean model and two more sophisticated variants, and we consistently observe a significant dropping trend in product diversity. These results make it possible to frame and start to test specific hypotheses for explaining the dropping product diversity trend.

\section{Introduction}

For decades economists have been using diversity to gauge the productivity and stability of regional economies, and this has motivated continuing efforts to craft better ways to measure diversity \cite{wagner1993measure,siegel1995,wagner2000regional,stirling2007general,van2019diversity}. The economic diversity of geographic regions has been correlated with higher levels of gross domestic product, and economic diversification is often promoted as a route to economic stability, growth and development \cite{dissart2003regional,Freire2017}. This paper focuses more narrowly on the diversity of the {\em products} bought and sold in the economy overall. The diversification of products produced by important individual firms has been studied \cite{tallman1996effects,anderson1999pricing}, and so has the diversity of products in markets with many kinds of firms selling many kinds of products at fluctuating prices to many kinds of consumers \cite{dixit1977monopolistic,yang1993monopolistic,d1996dixit}. Taking advantage of the existence of high quality public textual data, this paper focuses on the products of large US firms over the past two decades.

\begin{figure}
    \centering
    \includegraphics[width=0.7\linewidth]{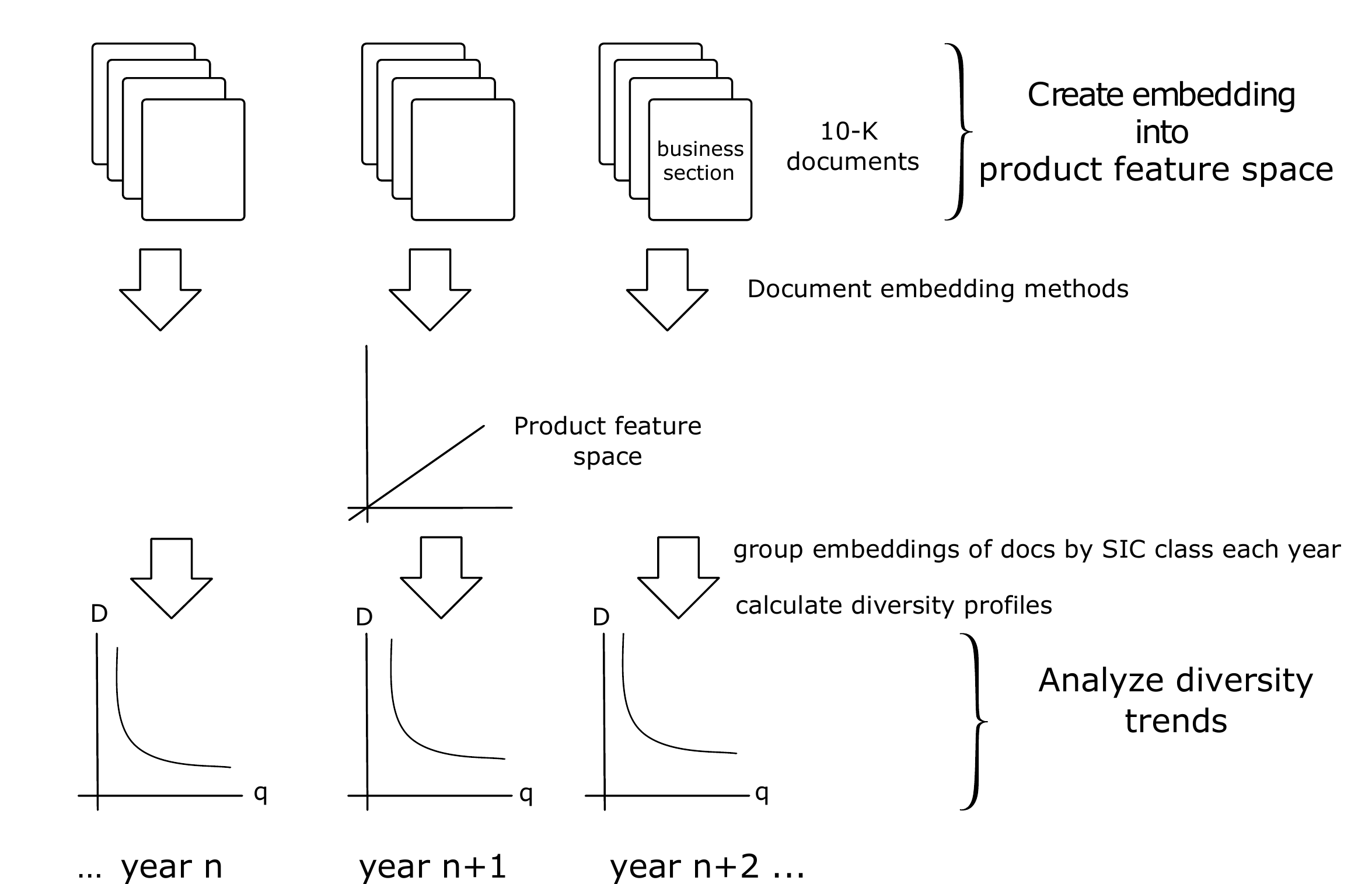}
    \vspace{0.1in}
    \caption{The workflow for a text-based analysis of product diversity.}
    \label{workflow}
\end{figure}

Some discussions of product diversity are theoretical and focus on the mathematical consequences of simple economic scenarios, but our focus is empirical and data-driven, and relatively theory neutral and free of economic assumptions. We simply observe the changing product diversity of large US firms, evident in their annual product descriptions, and describe the trends we observe. \added[id=AN]{In recent years, there have been similar efforts to draw ideas from quantitative biology, systems science and data mining to study the diversity of systems in social science \cite{bu-topic-2021, bollen-online-2018}.} \added[id=MB]{In economics, many papers design and apply standard indices of economic diversity and complexity (e.g,, \cite{mealy2019interpreting}), but atemporal data blinds us to how the indices have changed. The temporal data binning used here reveals how economic diversity and complexity have changed over the past generation and are trending today.} \added[id=AN]{By doing so, } our results are precise and quantitative. In addition, our methods are easily reproducible. We first embed annual documents describing each firm's products in a high-dimensional vector space, producing a model of the similarities among the products of large US firms. As shown in Figure~\ref{workflow}, we then group the vectors by SIC class to obtain product-focused vector representations for industry classes. The diversity of those products is calculated from this classification for each year. We focus on three different document embeddings: a Boolean embedding modeled after the current industry standard in product-focused industry classification  \cite{hoberg2010product,hoberg2016text}, a slightly more sophisticated TF-IDF embedding, and a more complex Paragraph-Vector  Distributed Memory (PV-DM) embedding. All of the models are first evaluated by measurement of their Industry Specificity relative to the Standard Industrial Classification (SIC) and evaluation of the {\em a priori} plausibility of their firm clusters. Models that pass these tests are each used to measure the diversity of the products of large US firms over the past two decades. In order to identify diversity trends that are robust, we employ a suite of more or less complex ways to measure diversity, including a baseline measurement based merely on each firm's SIC classification. This enables us to identify diversity trends that are robust across a variety of models.

By doing so, we provide evidence during 1997-2017 of a falling trend in the diversity of products offered by large US companies. This evidence comes from a consensus of semantic-vector models trained on a corpus of 10-K documents from 1995-2019 that describe the products of those firms. This trend is further corroborated by the text-free model based just on SIC Codes. We conclude by evaluating a number of hypotheses for how to explain the trend of dropping diversity.

Our work is one of a growing number of text-based analyses of economic topics, such as banking, finance, accounting, mergers and acquisitions, or corporate innovation and fraud. Many use topic modeling methods akin to our methods \cite{hoberg2014product,hoberg2015redefining,hoberg2017fraudulent,bellstam2020text} and apply them like we do to 10-K documents \cite{hoberg2014product,hoberg2015redefining,hoberg2017fraudulent}, while others mine other kinds of documents, such as IPO prospectuses \cite{hanley2010information,agarwal2017public,lowry2020information} and analysts' reports and regulatory filings \cite{popadak2013corporate,israelsen2016does,agarwal2017public,goldsmith2016parsing}.

\added[id=MB]{Our work also reflects the expanding diversity of applications of NLP and machine learning methods. Bergeaud and colleagues \cite{bergeaud2017classifying} used a similar methodology to classify patents by training models on patent documents, and their success motivates our application to economic documents of a more sophisticated methodology that was recently used to visualize and quantify the open-endedness of the evolution of technology \cite{bedau2019open}. Very similar computational methods were also recently used to infer models of the periodic table of elements from a training corpus of chemical documents \cite{tshitoyan2019unsupervised}. These methods have even successfully predicted the biological function of a protein from its amino acid sequence, by training models on a huge corpus of amino acid sequences \cite{Rives, nambiar2020transforming, Chris}---a vivid demonstration of the power and generality of our methods.}

\section{SIC model of firm similarity}

Many measures of diversity operate on kinds or classes of things. Firms are regularly classified into two hierarchical classifications: The Standard Industrial Classification (SIC) \cite{SIC} used by the Security and Exchange Commission (SEC) and the more recent North American Industry Classification System (NAICS) \cite{NAICS}. Both classifications were manually designed by experts and are updated by hand as industries evolve. In general, the NAICS classifies companies according to the processes by which they produce products, while the SIC classifies them according to the types of products they produce \cite{NAICS}. Given our present purpose of measuring diversity of products, this paper uses the SIC classification of firms when measuring the diversity of their products. 

\begin{figure}
\includegraphics[width=0.95\linewidth]{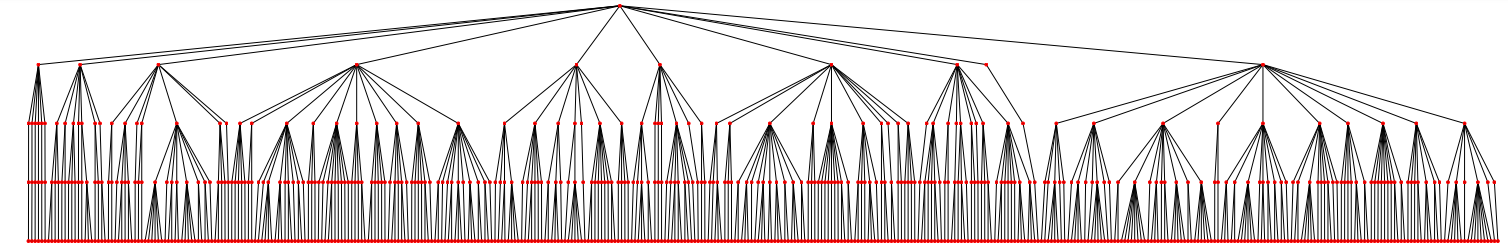}
    \caption{The SIC hierarchical classification tree of the documents in the training corpus, with Divisions (top level), Major Groups (second level), Industry Groups (third level), and Codes (bottom level).}
    \label{SIC4tree}
    \vspace{-.1in}
\end{figure}
 
In a hierarchical classification tree like the 4-digit SIC classification scheme, individual firms $i$ and $j$ are leaves at the bottom of a 4-level branching tree structure. For example, the SIC hierarchical classification tree depicted in Figure~\ref{SIC4tree} has 10 Divisions at the top level right below the tree's root, 83 Major Groups at the second level, 248 Industry Groups at the third level, 399 Codes at the forth level; each individual firm is classified and under exactly one SIC Code. Each large US firm has a 4-digit SIC Code that specifies the firm's Division, Major Group, Industry Group, and Code.

\begin{figure}
\centering
    \includegraphics[width=.8\linewidth]{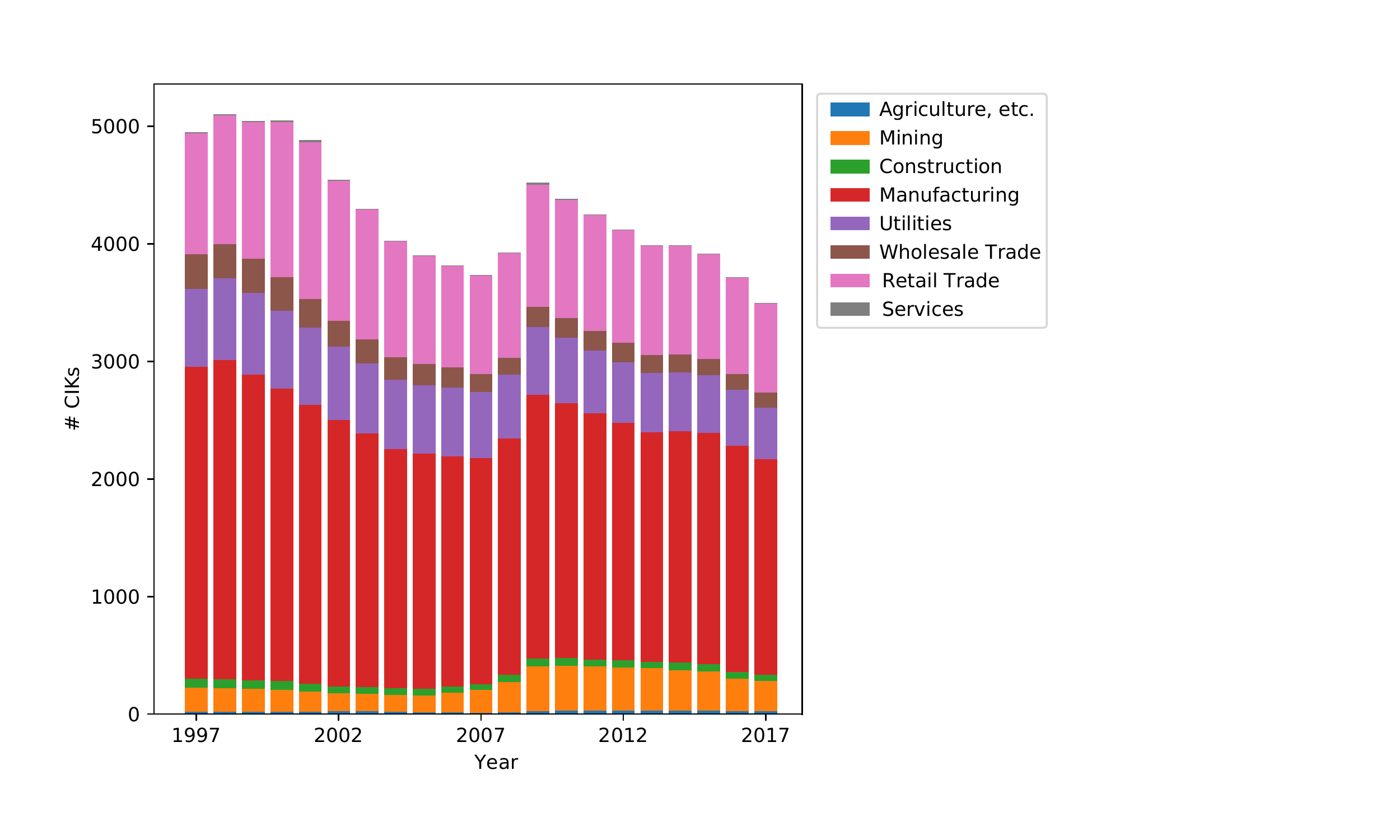}
    \caption{Annual count of 10-K documents in the training corpus; colors indicate SIC Divisions.
    }
    \label{firmcounts}
	 \vspace{-.1in}
\end{figure}

Figure~\ref{firmcounts} shows the number of large US firms (in our training corpus) over the past twenty years. We can track changes in the number of firms in each SIC Division because firms in the Figure are colored by their SIC Division. We can see that most SIC Divisions have shrunk somewhat so far in this century, although SIC Divisions typically retain roughly the same fraction of firms. The size of the Manufacturing Division dominates the pool, followed by Retail Trade and Utilities. The smallest Divisions are Services (almost invisible at the bottom of the bars), Construction, and Wholesale Trade. The only Division that shows significant growth in the past twenty years is Mining, which ended much larger than it started.

A simple gauge of the similarity of two firms is their {\em distance} from one another in the four-level SIC classification tree. We define the {\em distance} between firms $i$ and $j$ as the length of the shortest tree walk (sequence of adjacent nodes) between leaves $i$ and $j$. The number of sub-classes in the SIC classification tree varies significantly across the different nodes in the tree. To create more distance between firms classified under especially heavily branching nodes, we define the length of a walk as the number of SIC codes that fall under its highest node. 

\begin{figure}
\centering
    \includegraphics[width=0.6\linewidth]{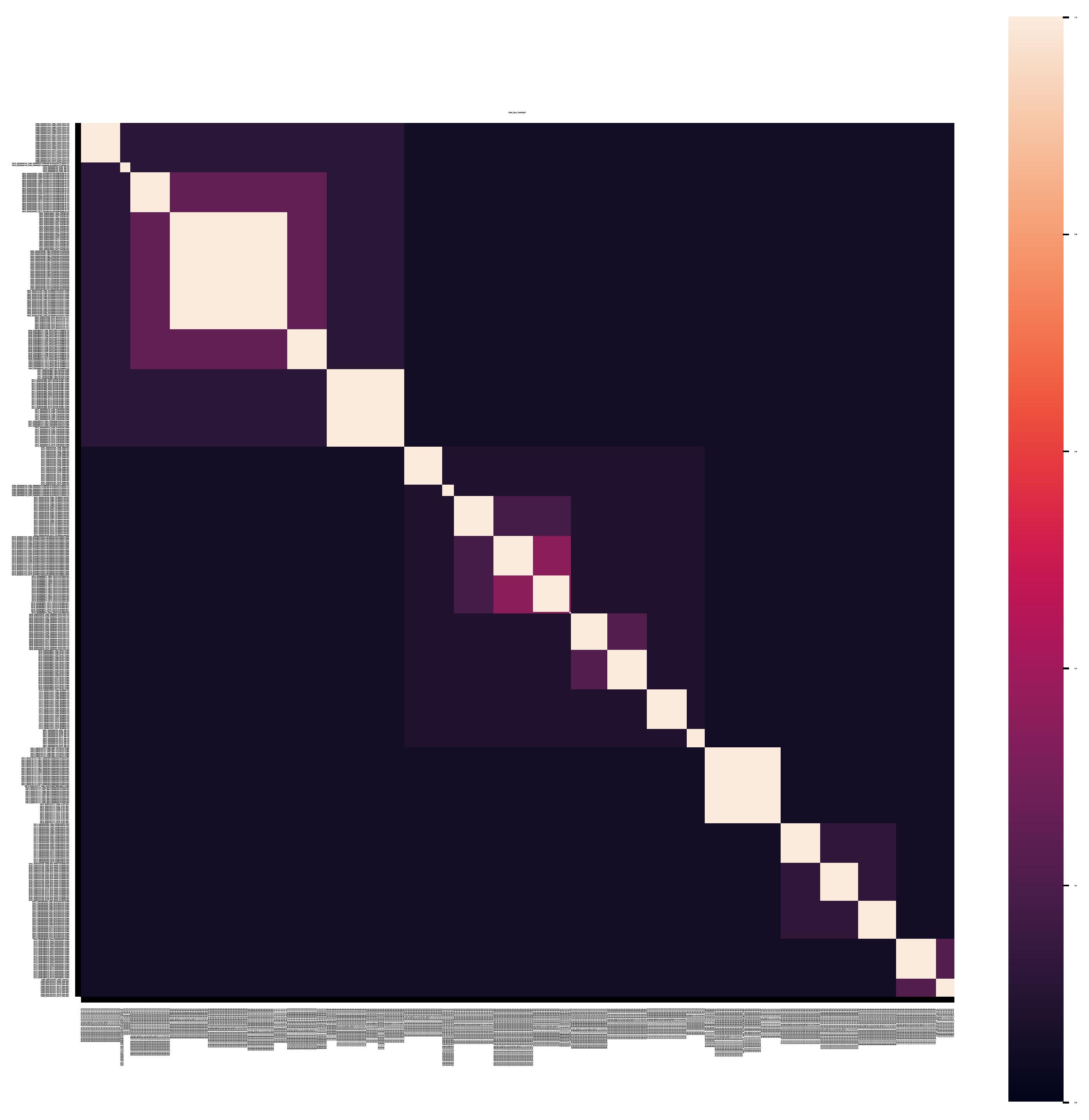}
    \caption{Heatmap of the similarity matrix among twenty years of products products by the firms in the DJIA. Rows (columns) of firms are ordered first by their 4-digit SIC Codes, and then by the CIK of each firm. The white squares along the diagonal show that pairs of documents with the same SIC Code are interpreted as having perfect similarity, and the fainter rectangles falling away from the white squares show that pairs of firms from coarser-grained SIC groups have a uniform intermediate degree of similarity.
    }
    \label{heatSIC4}
\end{figure}

The Standard Industrial Classification (SIC) tree has been carefully designed by human experts; it has passed the test of time and is widely used. We use it here to define a simple trustworthy metric of firm similarity against which to compare more sophisticated alternatives. This firm similarity metric based on proximity in the SIC classification tree is a crude representation of the similarities of actual firms. For example, the SIC tree proximity metric assigns a perfect similarity to every pair of firms in the same SIC Code, and it assigns identical similarities to all pairs of firms connected through the same highest node. This metric has a perfectly simple and predictable form, consisting of a number of rectangular fields with absolutely uniform similarity (Figure~\ref{heatSIC4}).

Embedding firms in semantic vector spaces provides a much more sensitive and product-centric measure of firm similarity. Each individual firm has a unique location in the vector space, which yields a fine-grained measure of the similarity of each pair of firms. The current industry standard in precise firm similarity matrices for large US firms is a simple Boolean word-vector embedding of documents \cite{hoberg2010product,hoberg2016text}. We construct and study an analogous Boolean word-vector model of product similarities, and we also construct and study two more sophisticated vector spaces. After confirming the plausibility of all of the models, we examine what they reveal about trends in the diversity of products of large US firms.

\section{Semantic vector model-training corpus}
\label{10k_des}

\begin{figure}
	       \centering
    \includegraphics[width=1.0\linewidth]{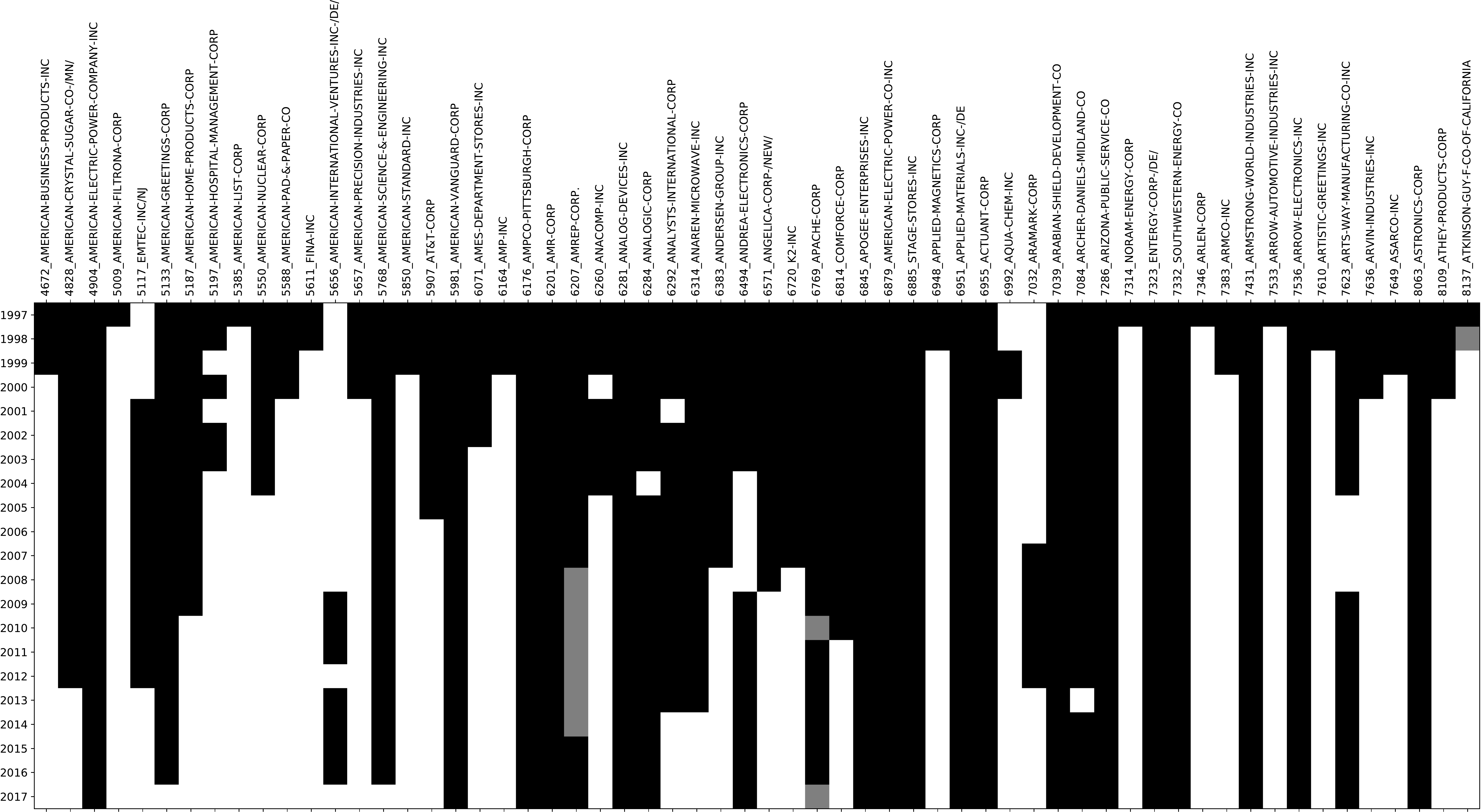}

    \caption{Map of annual documents for 30 firms in the model training corpus. Documents are sorted into firm (columns) and filing year (row). Black cells indicate documents in the corpus, grey cells indicate documents in the SRAF dataset that were removed because they lacked the information we needed. White cells indicate years when there is no document for that company in the SRAF dataset. }
	 \label{existencemap}
\end{figure}

In order to build the product vector space we use the Form 10-K, a document filed with the SEC by any company with more than $\$10$ million in assets with ownership by $2000$ or more individuals. The 10-K filing ``provides a comprehensive overview of the company's business and financial condition'' \cite{SEC:form-10k}. Companies that file 10-K forms with the SEC are large US firms. Taken together the 10-K corpus is a complete, accurate, standardized, publicly available annual description of the products produced by every large US firm, and it was used to train the current industry standard in quantitative firm similarity measurement \cite{hoberg2010product,hoberg2016text}.

We use the section of 10-K documents typically labeled ``Part 1 Section 1: Business''. The Business section of a firm's 10-K describes significant products the firm offers to their customers, what markets the firm operates in, and any subsidiaries it owns \cite{hoberg2010product,hoberg2016text}. If it exists, we exclude the part of the Business section typically labeled ``Section 1A: Risk Factors\new{,}'' leaving only details relevant to offered products.

We obtain 10-K, 10-K405, and 10-KSB documents from 1993 through 2018 from the Software Repository for Accounting and Finance (SRAF) \cite{Loughran2016sraf}. The 10-K documents do not all have one standardized format, and their heterogeneity makes it a challenge to extract exactly their Business sections. SRAF stage-one parsing removes various markup from the documents and removes tables. 
Figure \ref{firmcounts} shows the number of unique companies which file for each year in our dataset (broken down by SIC division). 

After obtaining the data we extract the desired section by way of a series of regular expressions designed to catch common formats as well as a more flexible keyword based program. In total approximately 12\% of documents cannot be parsed by either the regular expression or the keyword approaches. As Figure \ref{existencemap} illustrates, filing data for each company exists for only a subset of the years considered, but in general our programs 
are able to extract business sections from filings whenever the filings exist.

We evaluated the success of our extraction by manually checking both the extracted business 
sections to ensure that they were complete and did not contain extra text, as well as by reading through the unparseable documents to see if there was actually business section information lost by excluding those filings. Analysis of 50 randomly chosen extracted business sections revealed 49 of them to be correctly pulled from the corresponding 10-K forms. The errant filing was such that sections beyond the business section were included in the extracted text. Manual analysis of 100 randomly chosen unparseable filings found that 90 of them contained no business section at all, while the other 10 had either especially non-standard formatting, extremely short business sections of less than 1000 characters, or combined their business and properties sections into a single section which made the relevant details of the section hard to distinguish from the irrelevant details. These analyses make us confident that we are building models on a dataset which is reasonably complete as well as textually relevant.

Once the appropriate sections are extracted they are preprocessed to only include nouns as suggested by \cite{hoberg2016text}. In addition, we convert all alphabets to lower-case, remove extra white spaces, numeric values, stop words and words shorter than 3 characters long. In order to faciliate comparison with \cite{hoberg2016text}, we also remove from the training corpus any filings which do not have Compustat Global Company Keys, which lack at least a year of lagged Compustat data, or which are financial firms (SIC Codes 6000-6999), again following \cite{hoberg2016text}. While the notion of a product can be extended to include some of the things that are ``produced'' by some financial firms, many large US financial firms do not offer the consumer products on which our analysis focuses. This last step reduces the number of individual documents in our training corpus from 179,717 to 107,500. 

The number of different CIKs in the 10-K documents filed each year is plotted in Figure~\ref{firmcounts}. This plot shows the size and Divisional composition of the documents used to train semantic vector models. The figure shows that across the years some SIC Divisions consistently contain at least an order of magnitude more 10-K forms than some other Divisions, with Manufacturing always dominating and three other Divisions (Agriculture etc., Construction, and Wholesale Trade) always lagging far behind. The figure also shows a largely consistent downward trend in the number of firms, with an exceptional uptick during 2008 and 2009.

\begin{figure}
\centering
    \includegraphics[width=0.45\linewidth]{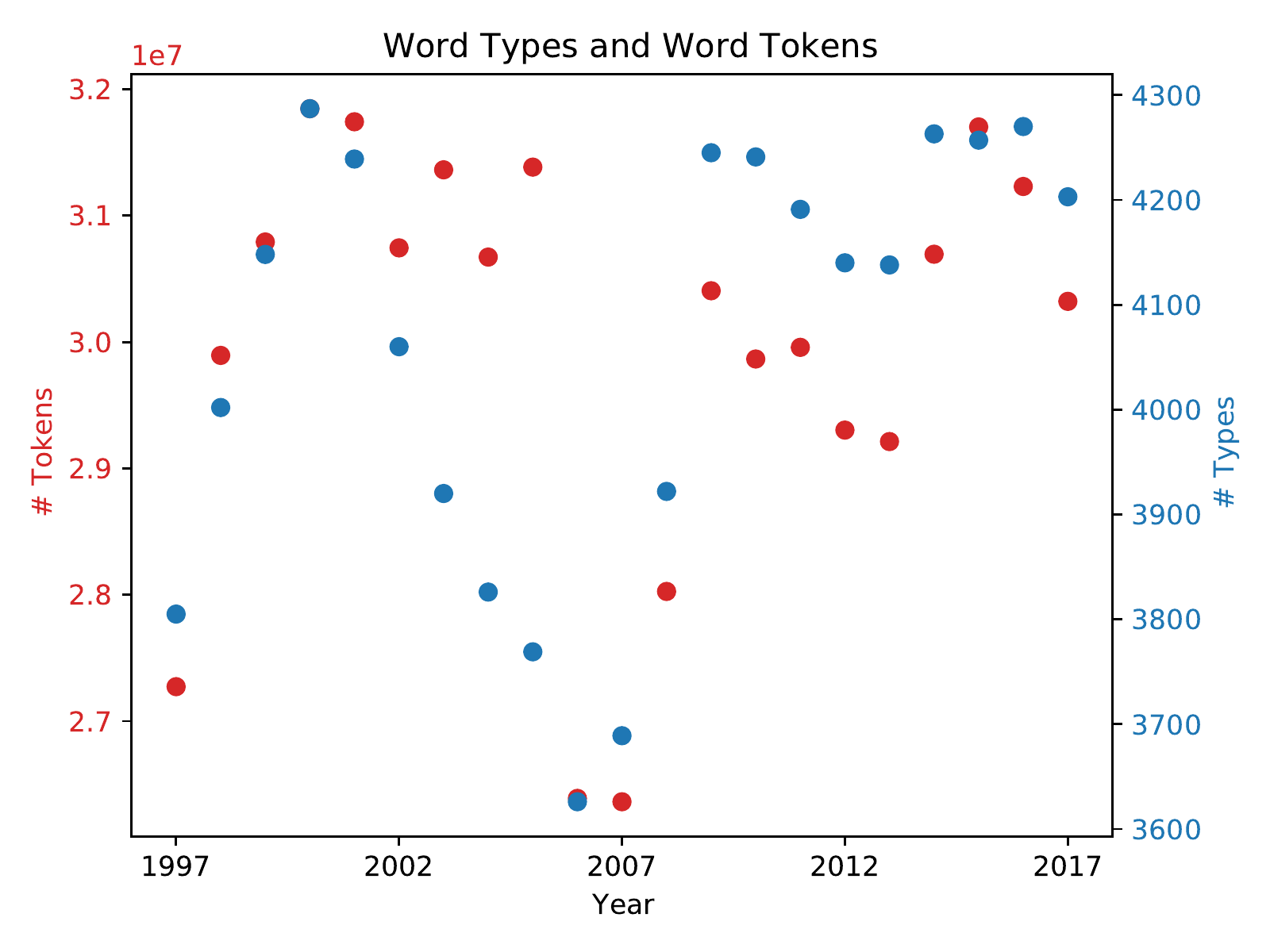}
    \includegraphics[width=0.45\linewidth]{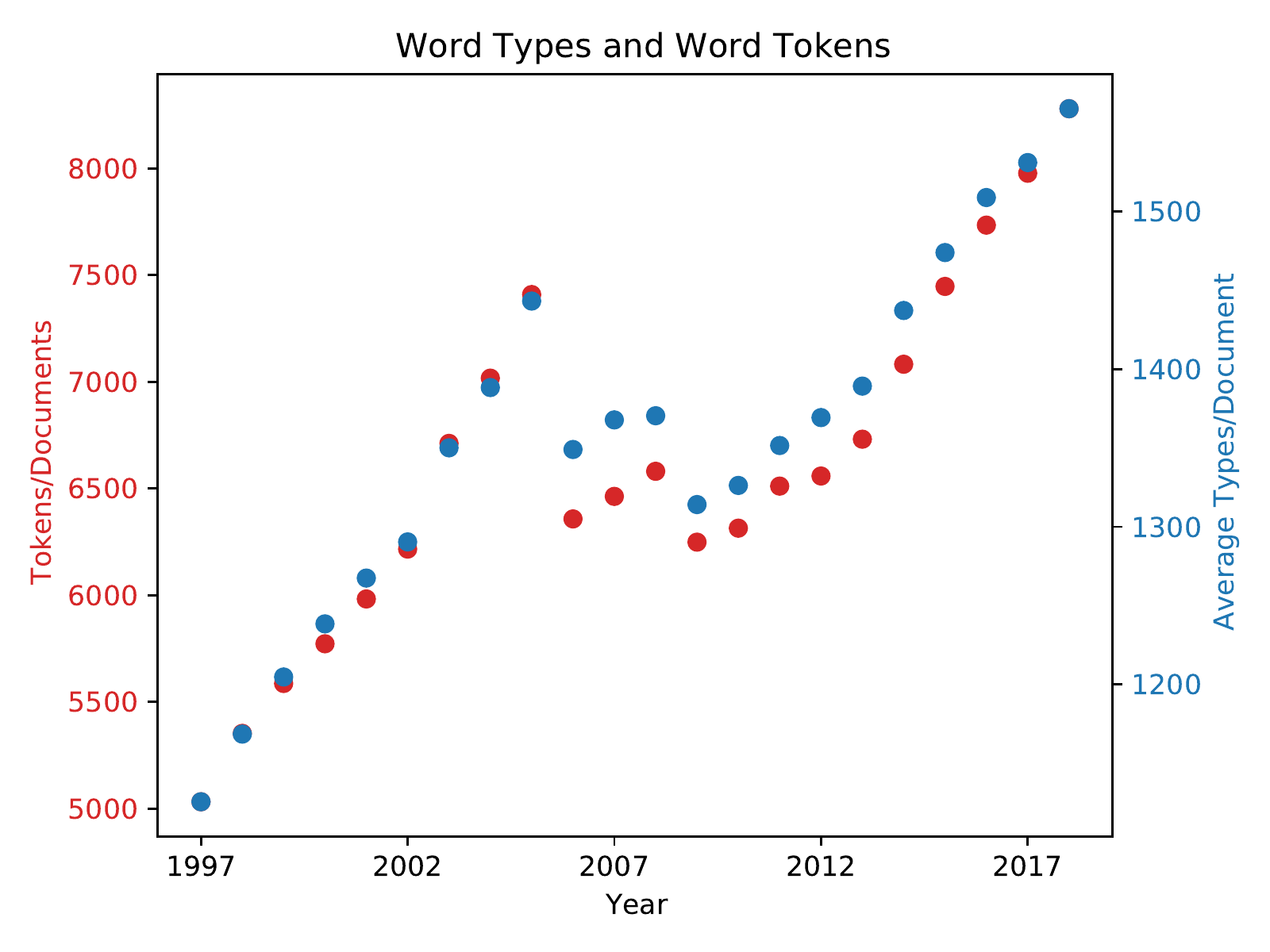}
    \caption{Left: The number of word tokens and word types in annual batches of 10-K documents in training corpus, during the first two decades of this century. Right: The mean number of word tokens and types in each document in each batch, over the same years.
    }
    \label{tokensandtypes}
\end{figure}

Text documents contain a number of {\em types} of words, and each type of word might have many different instances or {\em tokens} in a document. The left of Figure~\ref{tokensandtypes} shows the total number of word tokens and types in the training corpus each year, and the right of the figure shows the average number of word tokens and types in each document each year. Aside from a drop during 2006 and 2007, the total number of word tokens and types are each fairly consistent overage. By contrast, the {\em mean} number of word tokens and types per document rose significantly. (The mean also shows an analogous drop during 2006-2009.) The mean number of word tokens in a training document grew sixty percent during the first two decades of this century, and the number of word types grew more than forty percent.

\begin{figure}
\centering
    \includegraphics[width=0.32\linewidth]{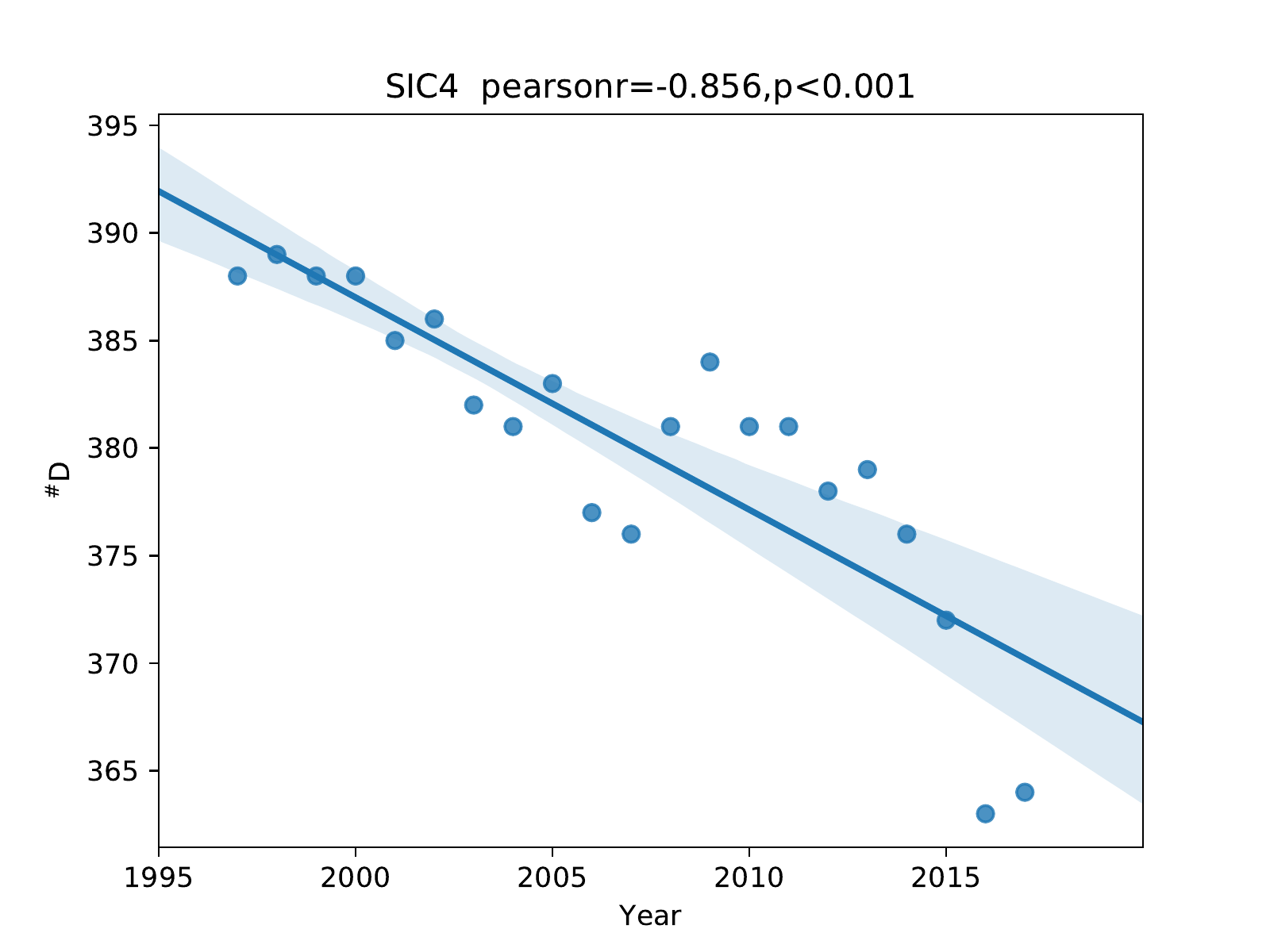}  
    \includegraphics[width=0.32\linewidth]{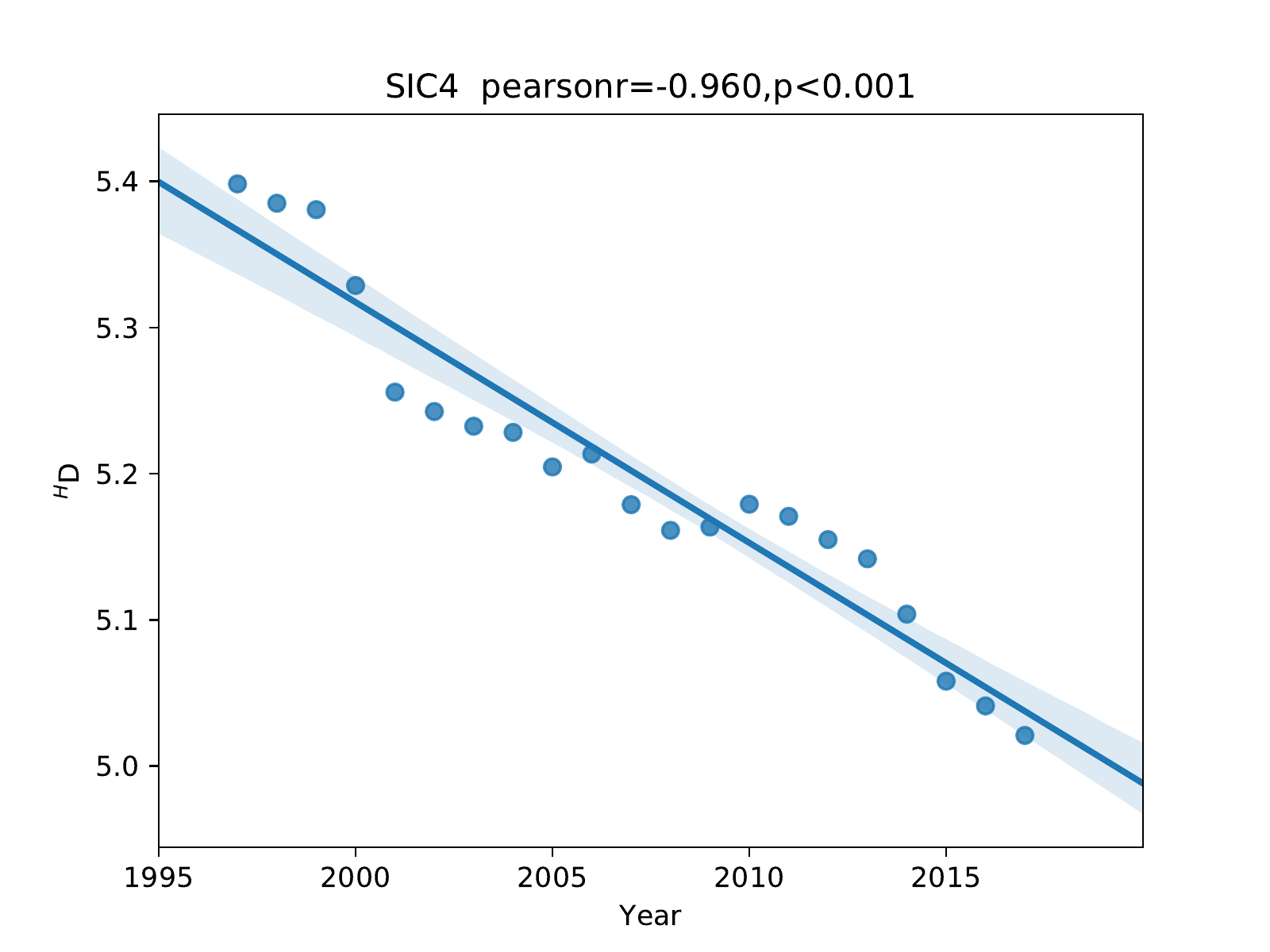}
    \includegraphics[width=0.32\linewidth]{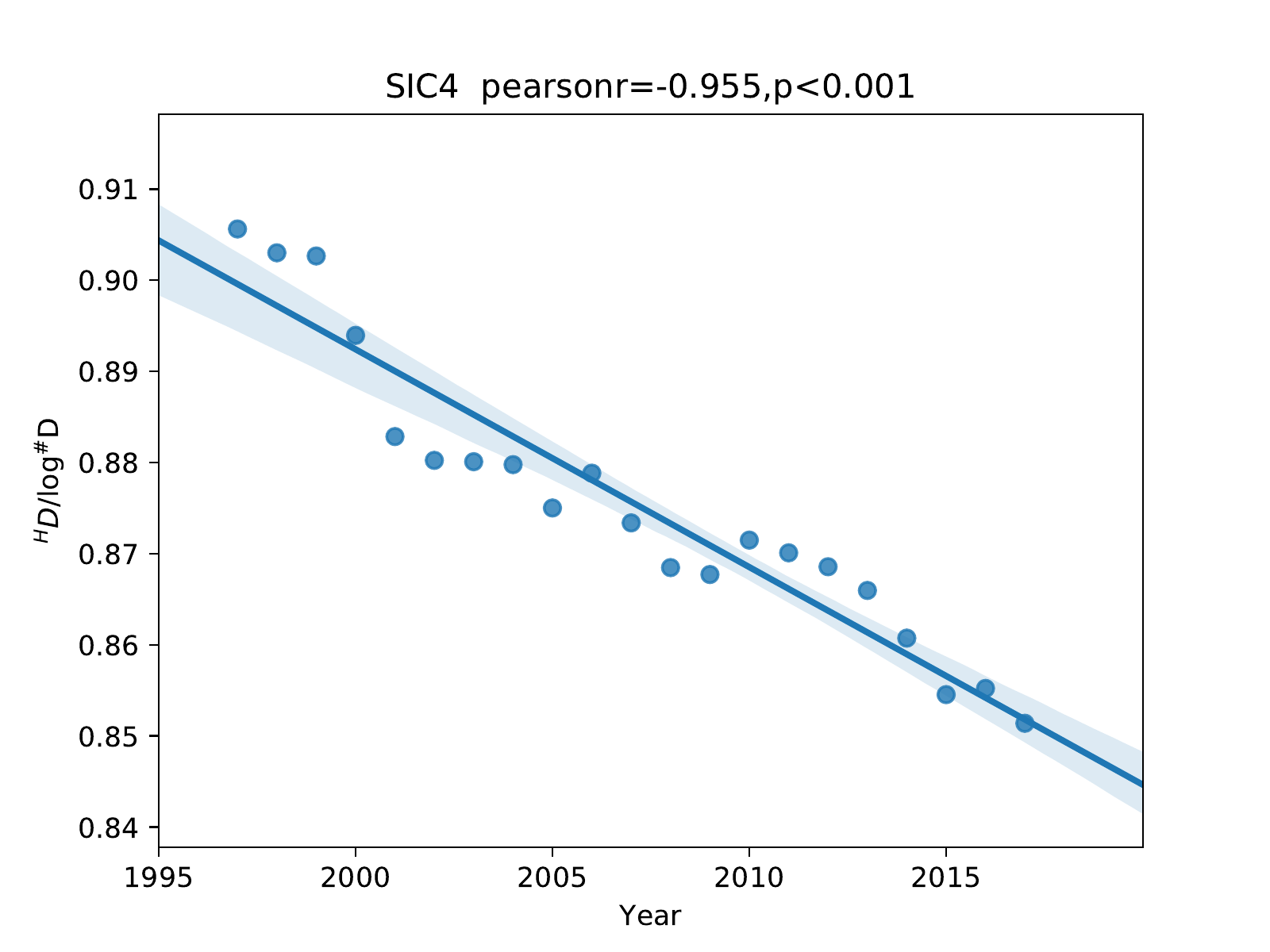}
    \caption{Left: The number of the 4-digit SIC Codes that are instantiated, during the first two decades of this century.
    Middle: Shannon entropy of the distribution of document counts per SIC Code during those years. 
    \added[id=MB]{Right: Normalized Shannon entropy of the same distribution during those years. Each plot includes a scatter plot of diversities and a linear regression fit with its 90\% confidence interval.}
    }
    \label{SIC4}
\end{figure}

Further analysis of the annual batches of 10-K documents in the training corpus shows that the number of instantiated SIC Codes has generally dropped during the first two decades of this century (Figure~\ref{SIC4}, left). At the start of the century an annual batch of 10-K documents covered firms from 392 SIC Codes, and twenty years later that number has dropped to 263 Codes. Figure~\ref{SIC4} (middle) shows twenty years of the Shannon entropy of the distribution of the number of instances of each SIC Code, revealing a general downward trend. \added[id=AN]{The entropy of a distribution reflects both its width (number of bins) and evenness (similarity of counts across all bins). Since the number of bins (instantiated SIC classes) varies by more than 7\% across the years we studied, it is also interesting to plot just the evenness of the distribution, which is shown by the normalized entropy:
\begin{align}
    ~^HD_{adj} &= \frac{~^HD}{\log ^\#D}
\end{align}
where $~^HD$ is the Shannon entropy and $^\#D$ is the number of bins (Figure~\ref{SIC4} (right)). The distribution's entropy and its width and evenness all display decreasing trends.}

The number of instantiated SIC Codes is a simple measure of the diversity of the products produced by large US firms, as are the normalized or non-normalized Shannon entropy of the distribution. But both product diversity metrics are crude, \added[id=AN]{because they ignore the different degrees of similarity between different SIC classes}.

\section{Embedding product descriptions with models}
 
 The documents in the training corpus are used to train a firm-similarity model that contains a vector representation of the products of each firm. Specifically, for every document $p \in F$, the embedding function is given by $f_e: p \rightarrow v_p$ where $v_p \in \mathbb{R}^d$. All these vectors are normalized to have a length of 1. Here we compare embedding methods for bag-of-word models and neural network models.

\subsection{Bag-of-words embeddings}
Bag-of-words models ignore the order of words in the training corpus and build vectors based just on the occurrence of the words. We study two different bag-of-words embeddings: Boolean  and Term Frequency-Inverse Document Frequency (TF-IDF).

In the Boolean  model, the vector for document $p$, $v_p$ is given by
\begin{align}
    v_p[i] = 
    \begin{cases}
    1,& \text{if } \Sigma[i] \in p\\
    0,              & \text{otherwise}
    \end{cases}
\end{align}
for every word $\Sigma[i]$ in the dictionary. Following \cite{hoberg2016text}, a word is included in the dictionary only if it appears in less than 20 percent of the documents in the training corpus. Removing very common words is important but it is arbitrary to set a threshold at precisely 20 percent. 

A more principled method is to replace the Boolean information about a word with the word's term frequency/inverse document frequency (TF-IDF) statistic---a commonly used measure of the relevance of each word in a document from a large corpus. The TF-IDF model defines $v_p$ with 
\begin{align}
    v_p[i] = 
    \begin{cases}
    count(\Sigma[i],p)*log(\frac{|F|}{docs(\Sigma[i],F)}), \\\qquad\text{if } \Sigma[i] \in p\\
    0,\text{~ ~ otherwise}
    \end{cases}
\end{align}
where $count(\Sigma[i],p)$ counts the number of times $\Sigma[i]$ occurs in document $p$, $docs(\Sigma[i],F)$ counts the number of documents $\Sigma[i]$ occurs in and $|F|$ is the total number of documents in the corpus \cite{idf}.

\subsection{Neural embeddings}
To obtain neural embeddings of firms in product space we use the Paragraph Vector Distributed Memory (PV-DM) model originally introduced in \cite{doc2vec}. For a window size of 3, given words $a_i, a_{i+1}$ and $a_{i+2}$ from document $p$, a neural network (NN) is trained to predict $a_{i+3}$. 
The PV-DM model then defines $v_p$, the vector representation of document $p$, as the $d$-dimensional hidden layer of the NN that is obtained from the document token. 
We omit details of the PV-DM architecture here since it is widely used and was followed here in all important respects. As suggested by the original paper \cite{doc2vec}, we set the number of dimensions of our PV-DM model at $d=300$, and we trained the model for 20 epochs with a starting learning rate of 0.025 (which decays linearly between epochs) and a window size of 8. To train our PV-DM model, we take advantage of the Gensim library developed by \cite{gensim}.

\section{Methods of analysis}



Before we use our models to make more sophisticated measurements of the diversity of the products, we first establish the plausibility of the embeddings of firms in semantic vector spaces produced by the Boolean, TF-IDF, and PV-DM models. We gauge model plausibility in two ways: One is to measure how much similarity the embeddings attribute to the firms within the same industries, where the industries are identified by some trusted source. The other is to examine whether the micro-structure of the embeddings fit with human common sense judgments of the similarity of well-known firms. 

\subsection{Industry specificity}

Existing classifications such as the SIC consider firms in the same industries to be relatively similar, and firms in distinct industries to be much less similar. The SIC is constructed by domain experts and is widely used by researchers and government offices, so it is safe to assume that each industry defined by a 4-digit SIC Code contains firms that are rather similar, much more similar than firms with different SIC Codes. So, one way to assess the plausibility of the vector embeddings of documents by individual firms is simply to check whether the average similarity of pairs of documents from firms in the same SIC Code is much higher than the average similarity of firms with different SIC Codes. The ratio of these two averages we term the {\em Industry Specificity (relative to the SIC)} of the similarity matrices produced by a given model. (See Appendix S1 for precise definitions.)

\subsection{Diversity}

{\em Diversity} of products is often measured in economics simply as the number of different types of commodities (goods, products) available in a marketplace \cite{dixit1977monopolistic,yang1993monopolistic,d1996dixit}. This approach is roughly analogous to the plot in of the number of different SIC Codes exemplified each year by large US firms (Figure~\ref{SIC4} left). Sometimes the distribution of types of commodities in a market is weighted in some way, such as by total sales, and diversity is then measured by something like the Shannon entropy of the distribution of products \cite{tallman1996effects,anderson1999pricing}, an approach analogous to the Shannon entropy of the distribution of SIC Codes instances shown in Figure~\ref{SIC4} (right). This entropy measure is quite simple, but is also rather crude, too crude, for example, to reflect the diversity of the firms within each SIC Code, or the ``distance'' between different SIC Codes within a given SIC Industry Group.

A more fine-grained approach is to measure the variance of the vectors in a product feature space by computing the number of dimensions needed to account for the bulk (here, 90\%) of the variance of all of the individual firm vectors in each year. This measure has the virtue of being built out of the local details of the embedding of firms in a product space, and the results are relative to that product space. This measure is easily applied to documents that have been embedded in any product space of interest, and here we use the Boolean, TF-IDF, and PV-DM vector spaces.

An even more fine-grained measure of the diversity of the products produced by a set of firms comes from a generalized measure of diversity from theoretical ecology. Once a classification with $s$ classes (as defined by four-digit SIC Codes) is obtained for a particular year $y$, a similarity matrix between different classes $Z_y$ and a normalized abundance vector $a = [a_{y,1},...,a_{y,s}]$ is calculated. The diversity is then defined as 
\begin{align}\label{DiversityEq}
    ~^qD(a_y,Z_y) = (\sum_i^s a_{y}[i] (\sum_j^s Z_{y}[i,j]a_{y}[j])^{q-1} )^\frac{1}{1-q}
\end{align}
where $q \neq 1$ is a sensitivity parameter \cite{sim_diversity} that controls how much the diversity measure emphasizes common versus rare industries. When $q$ is small, $~^qD(a,Z)$ gives as much importance to rare industries as common ones \cite{sim_diversity}; thus,  $~^0D(a,Z)$ is a measure of industry ``richness'' (the effective number of industries). By contrast, when $q$ is large, rare industries are de-emphasized and $~^qD(a,Z)$ includes information about the evenness of industries.

\section{Model plausibility results}%

We examine the firm-pair similarity matrices produced by the Boolean, TF-IDF, and PVDM models, and compare them for plausibility by comparison with the simple SIC model's similarity matrix (visible in Figure~\ref{heatSIC4}). Next, we test the plausibility of each model by seeing if they put similar firms in clusters, and if they give especially high similarity to pairs of firms with the same SIC Codes.

\subsection{Firm similarity matrices}

\begin{figure*}[h]
	       \centering
  \includegraphics[width=0.45\linewidth]{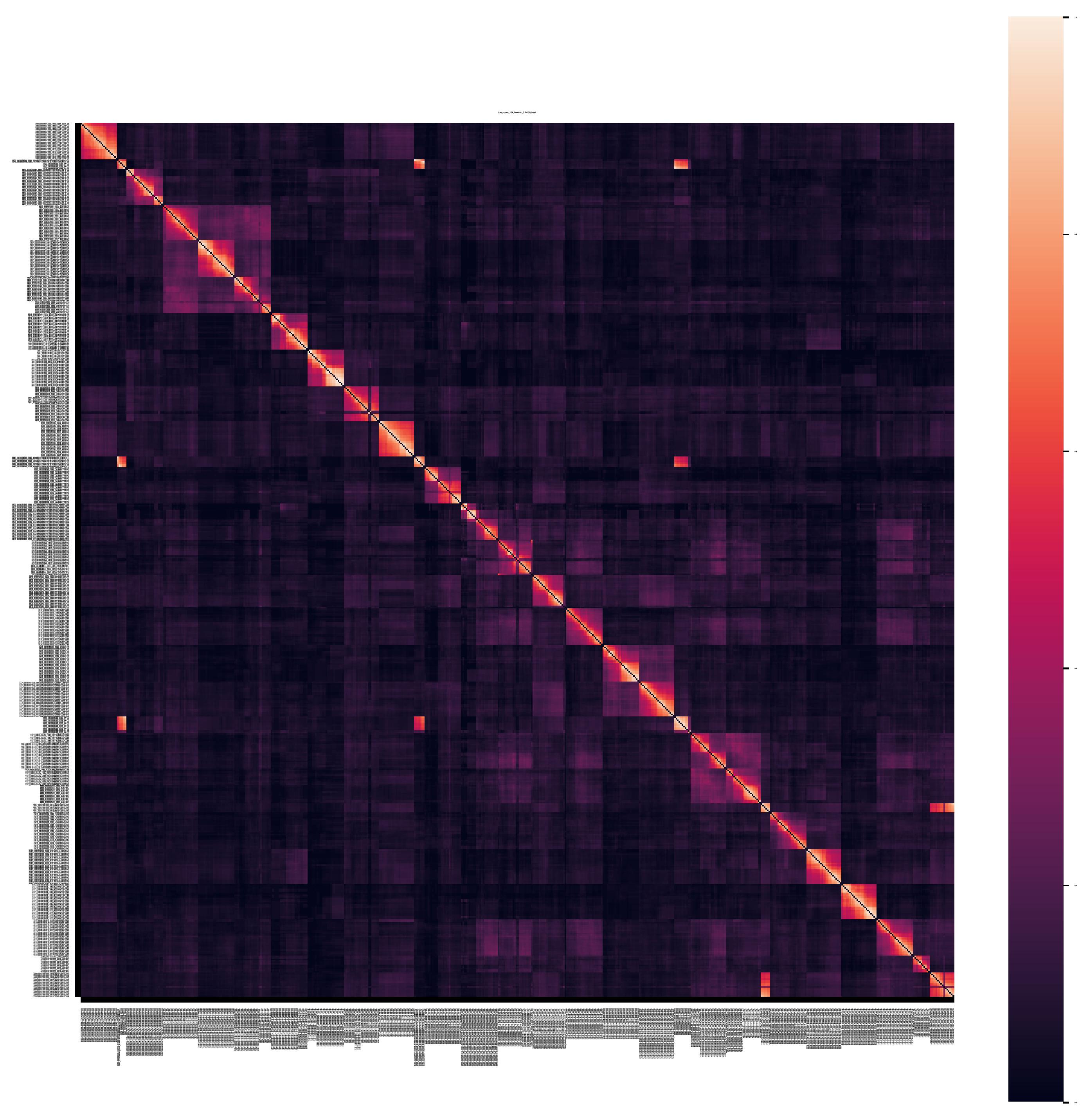}
  \includegraphics[width=0.45\linewidth]{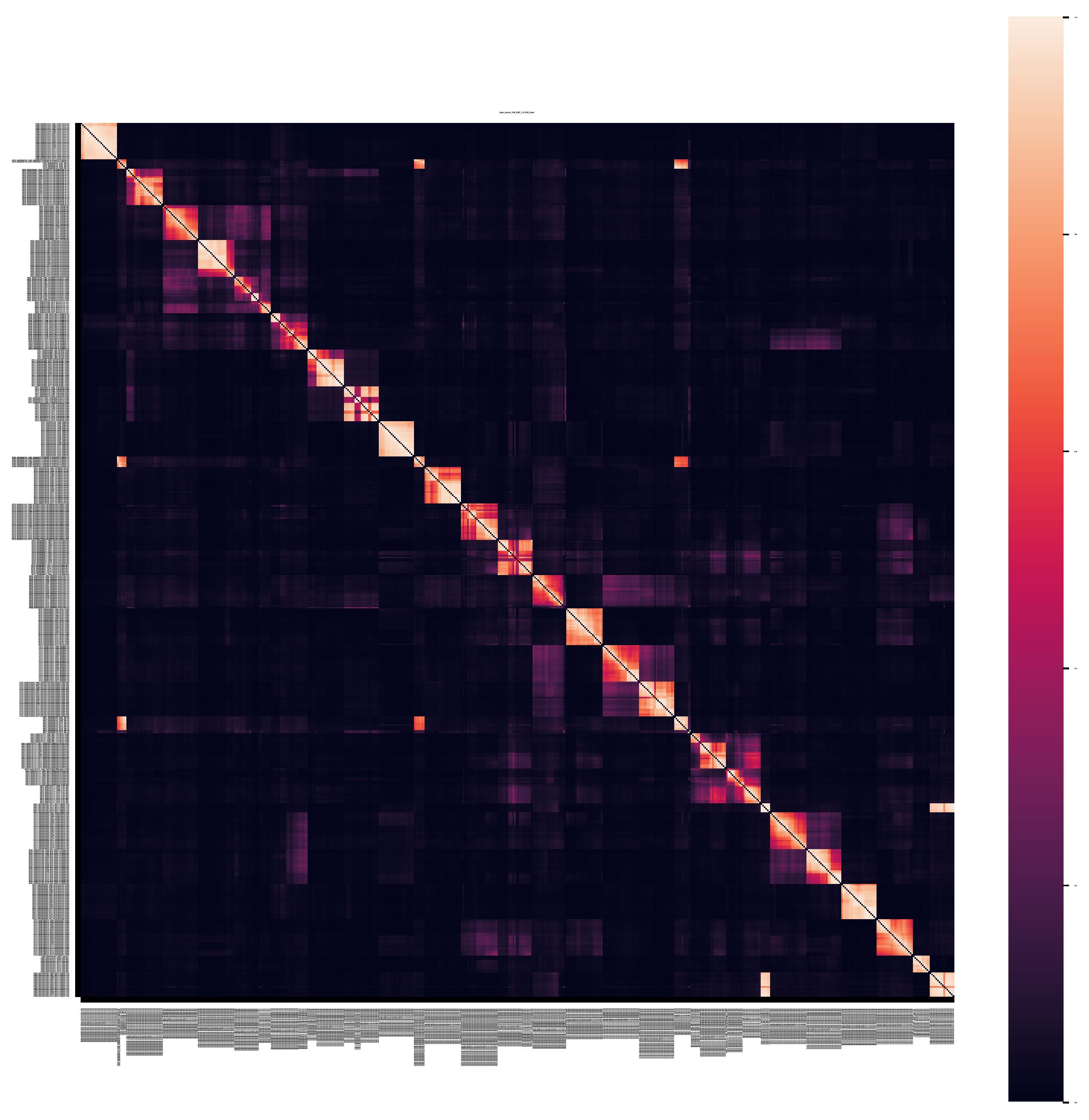}
  \includegraphics[width=0.45\linewidth]{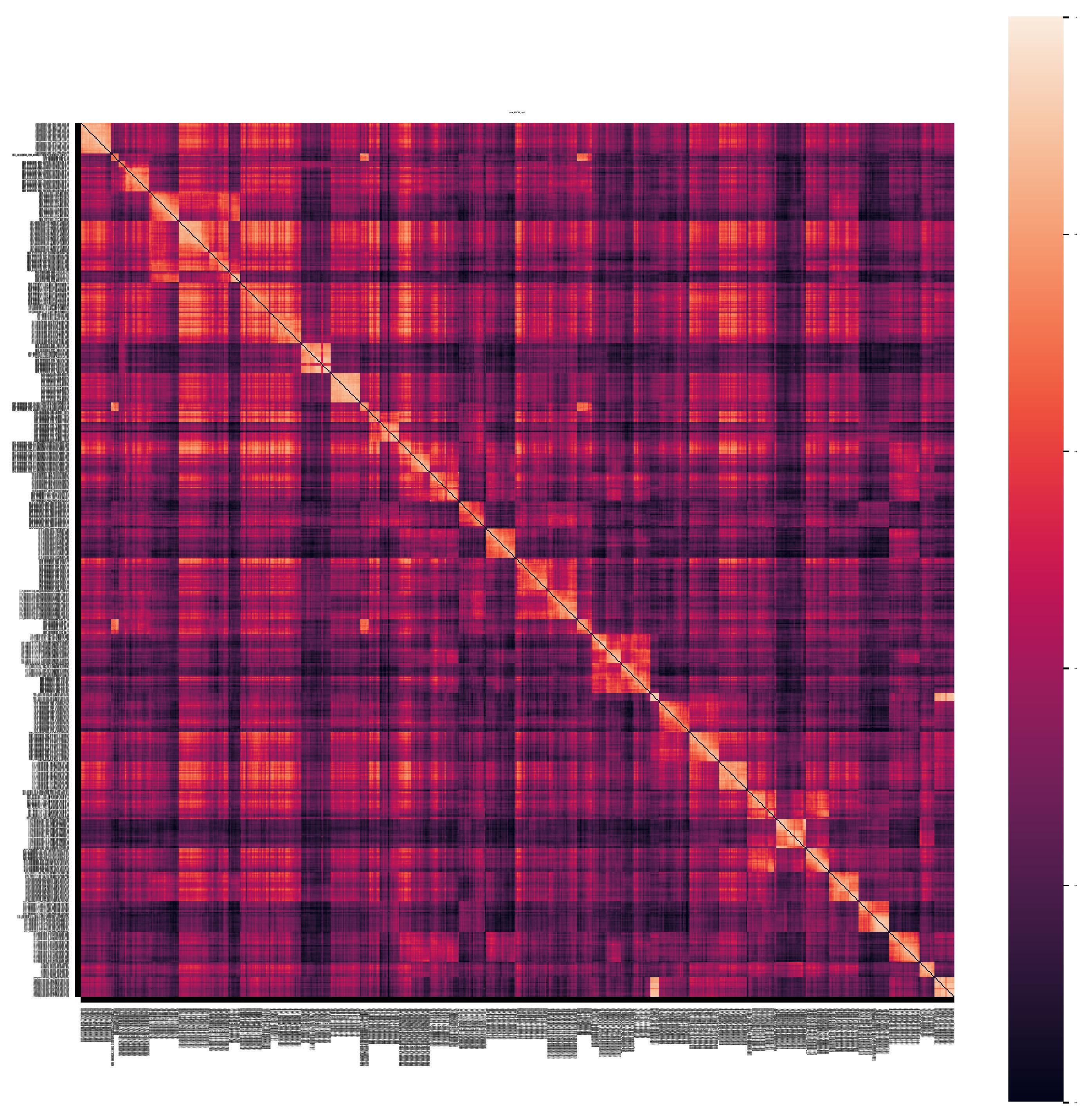}
    \caption{Heatmaps of firm similarity matrices of the products of the firms in the 2018 DJIA, according to the BOOLEAN (top left), TF-IDF (top right), and PVDM (bottom left) models. Compare with the SIC model similarity matrix shown in Figure~\ref{heatSIC4}.}
	 \label{heatmaps}
\end{figure*}

Figure \ref{heatmaps} shows heatmaps of the similarity matrices of twenty years of product descriptions by the firms listed in the 1997 Dow Jones Industrial Average (DJIA), according to the text-based (Boolean, TF-IDF, PVDM) models of firm level similarities.  All of the heatmaps show high similarity among firms with the same SIC Codes (squares of high heat along the diagonal), faintly reflecting the SIC model's white squares along the diagonal in Figure~\ref{heatSIC4}, and this pattern is quantitatively corroborated below with Industry Specificity measurements.

Furthermore, the firm similarity heatmaps shown in Figure~\ref{heatmaps} all show many differences in the similarity of different pairs of firms within the same SIC Code or higher-level SIC group. By contrast, the text-free SIC Code model depicts every pair of firms in each SIC group with the exact same level of similarity (Figure~\ref{heatSIC4}). The heatmaps contain a wealth of information about the different degrees of similarity attributed by each text-based model to each pair of individual firms.

\begin{figure*}[h]
	       \centering
    \includegraphics[width=0.45\linewidth]{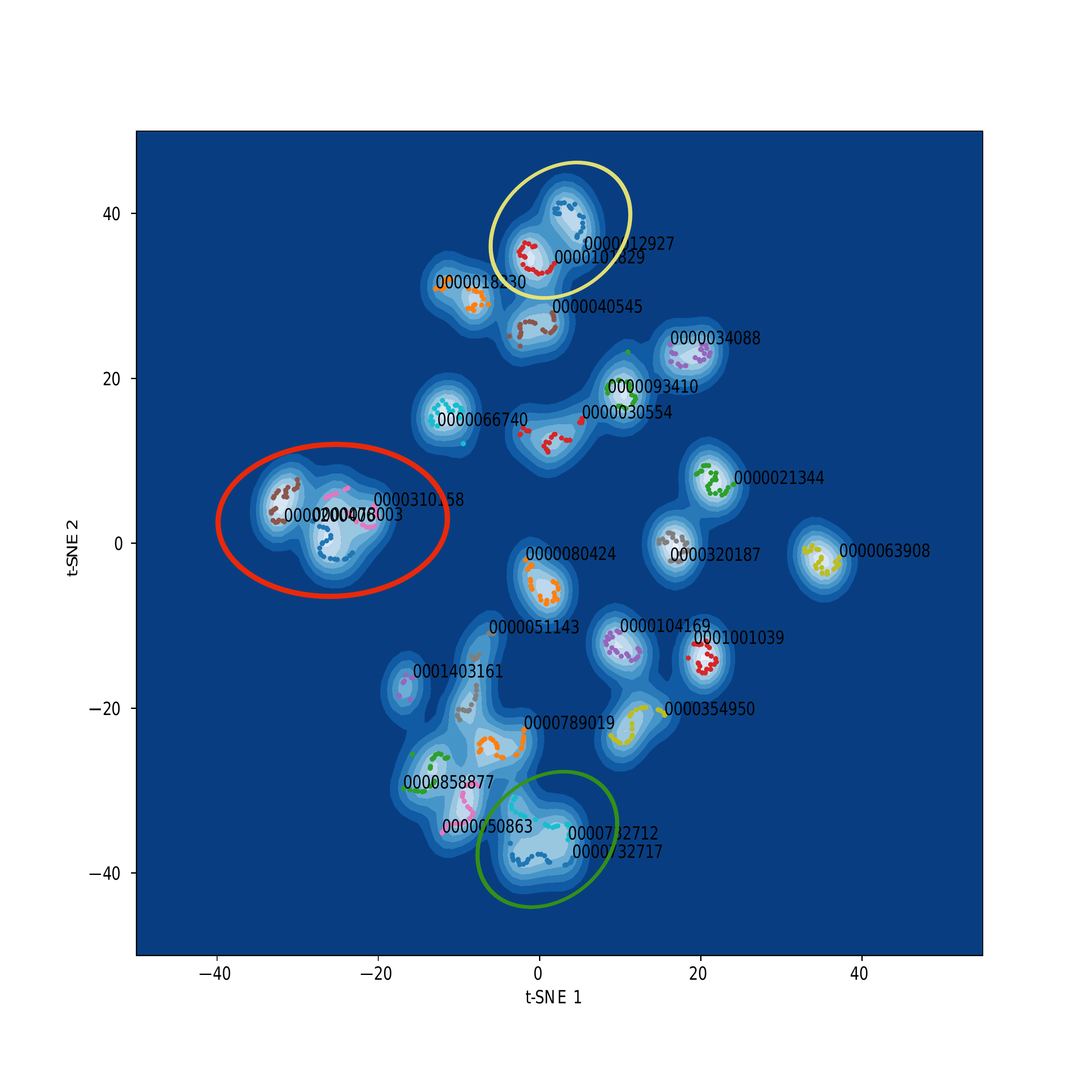}
  \includegraphics[width=0.45\linewidth]{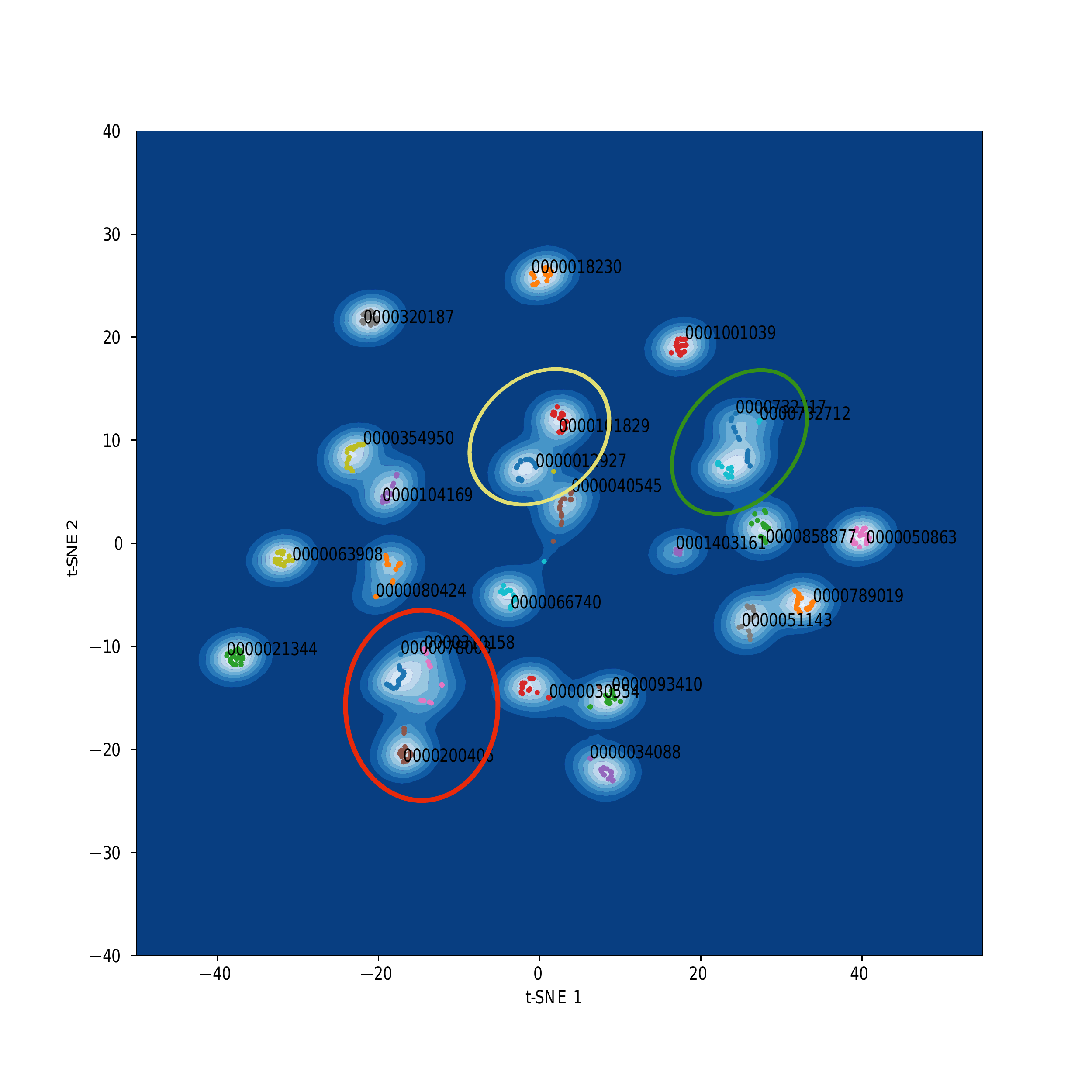}
  \includegraphics[width=0.45\linewidth]{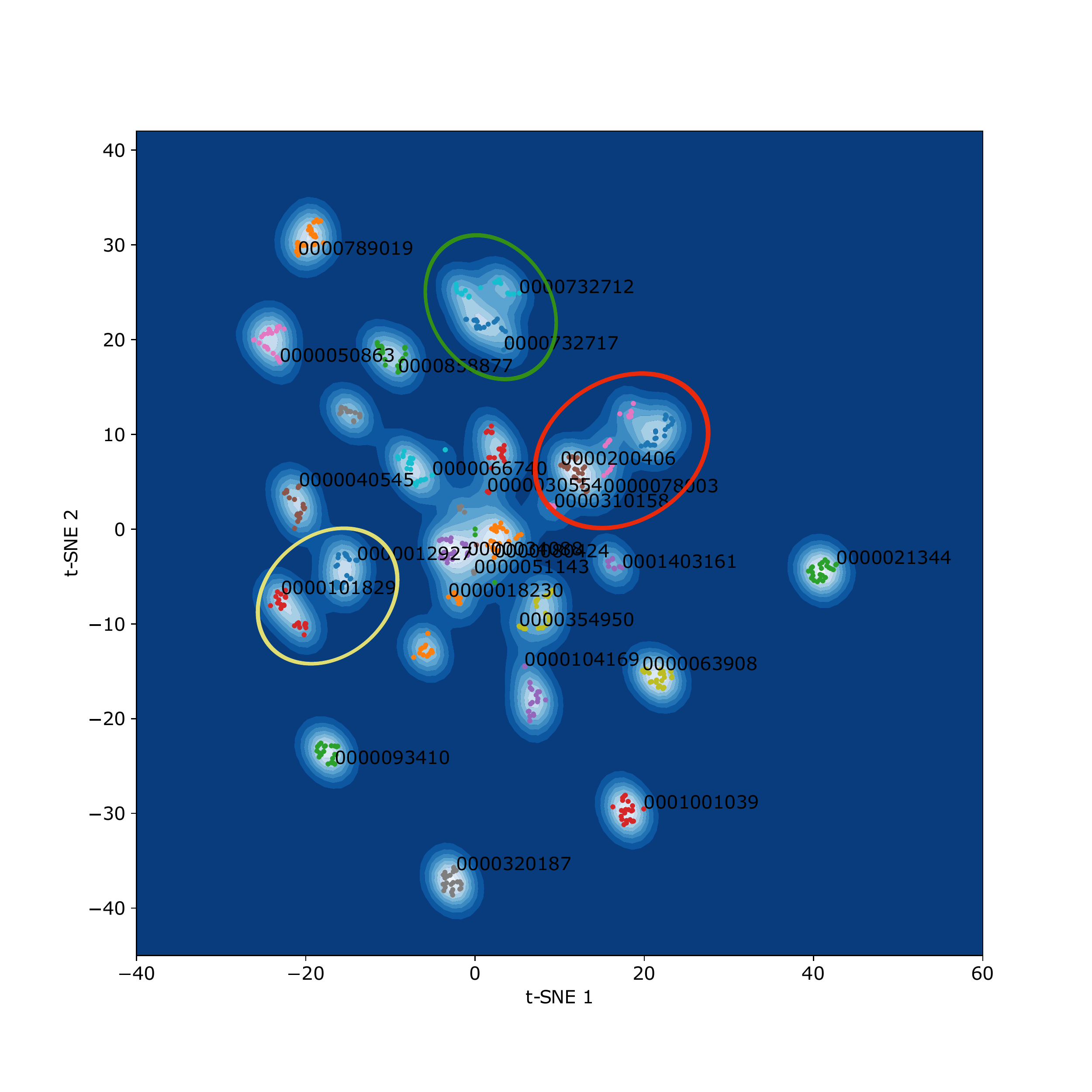}
    \caption{t-SNE of documents describing the products of twenty-five firms from the Dow Jones Industrial Average, embedded in the product spaces from models Boolean (left), TF-IDF (middle), and PV-DM (right). Points indicate the location of a given document, colored by firm; document clusters are labeled by dominant CIK. Firms that produce pharmaceuticals are circled in red, aerospace companies are circled in yellow and telecommunication companies are circled in green.}
	 \label{tsne}
\end{figure*}

\subsection{Micro-analysis of clusters}

To evaluate the text-based models and their consequent document embeddings, we focus on twenty five well-known firms: Boeing, Caterpillar, Cisco, Chevron, Disney, DuPont, General Electric, Home Depot, IBM, Intel, Johnson \& Johnson, Coca-Cola, McDonalds, 3M, Merck, Microsoft, Nike, Pfizer, Procter \& Gamble, AT\&T, United Technologies, Visa, Verizon, Wal-Mart, Exxon Mobil. All of these firms have been on the Dow Jones Industrial Average for much of this century. We evaluate the initial plausibility of each text-based model by seeing how well the proximity of each model's vectors for (documents about the products of) the twenty-five firms. 

We gauge the proximity of embedded documents in high-dimensional vector spaces using 2-dimensional t-SNE projections \cite{tsne} of the embedded documents. The t-SNE projections in Figure \ref{tsne} depict the location in product space of documents about our twenty-five reference firms, according to the Boolean (left), TF-IDF (middle), and PV-DM (right) models. Each point indicates the location of an individual document in a given year, and the dots are color-coded to by firm. CIKs label each document cluster.

A striking feature of Figure \ref{tsne} is that each cluster in the t-SNE projections contains documents from exactly one firm (one color); and documents about the products of different firms are in different clusters. This clear pattern is exactly what common-sense would expect from an embedding that reflects the real similarities among the products, and each each text-based model produces this pattern.

Common-sense also suggests that some firms make quite different products. For example, the products of the following firms are all relatively distinctive and different from each other: McDonalds (63908), Coca-Cola (21344), Nike (320187), Disney (1001039), Visa (1403161). 
Note that Figure~\ref{tsne} shows that the Boolean, TF-IDF, and PV-DM models all place McDonalds, Coka-Cola, Nike, Disney, and Visa in relatively isolated locations in their respective product spaces. 

Common sense also suggests that a plausible document embedding would put firms producing very similar products in nearby or overlapping clusters. For example, one would expect to see groups of nearby clusters (individual firms) containing the following groups of nearby DJIA firms (and their CIK number):
\begin{itemize}
    \item Johnson \& Johnson (200406), Merck (310658) and Pfiser (78003)
    \item Boeing (12927) and  United Technologies (101829) 
    \item AT\&T (732717) and Verizon (732712)
    \item Home Depot (354950) and Walmart (104169)
    \item Exxon Mobile (34088) and Chevron (93410)
    \item IBM (51143), Intel (50863), Cisco (858877) and Microsoft (789019)
\end{itemize}
Inspection of the t-SNEs in Figure \ref{tsne} confirms that the Boolean, TF-IDF, and PV-DM models all pass this additional test of common-sense. For ease of identification, the first three of these groups of firms are circled red, yellow and green.


This micro-analysis of the details of the embeddings of firms in the DJIA adds weight to the general plausibility of all three document embeddings studied here. The Boolean, TF-IDF, and PV-MD models all demonstrate a significant degree of common-sense realism and plausibility.

\subsection{SIC Industry Specificity results}

\begin{figure}
	       \centering
  \includegraphics[width=0.4\linewidth]{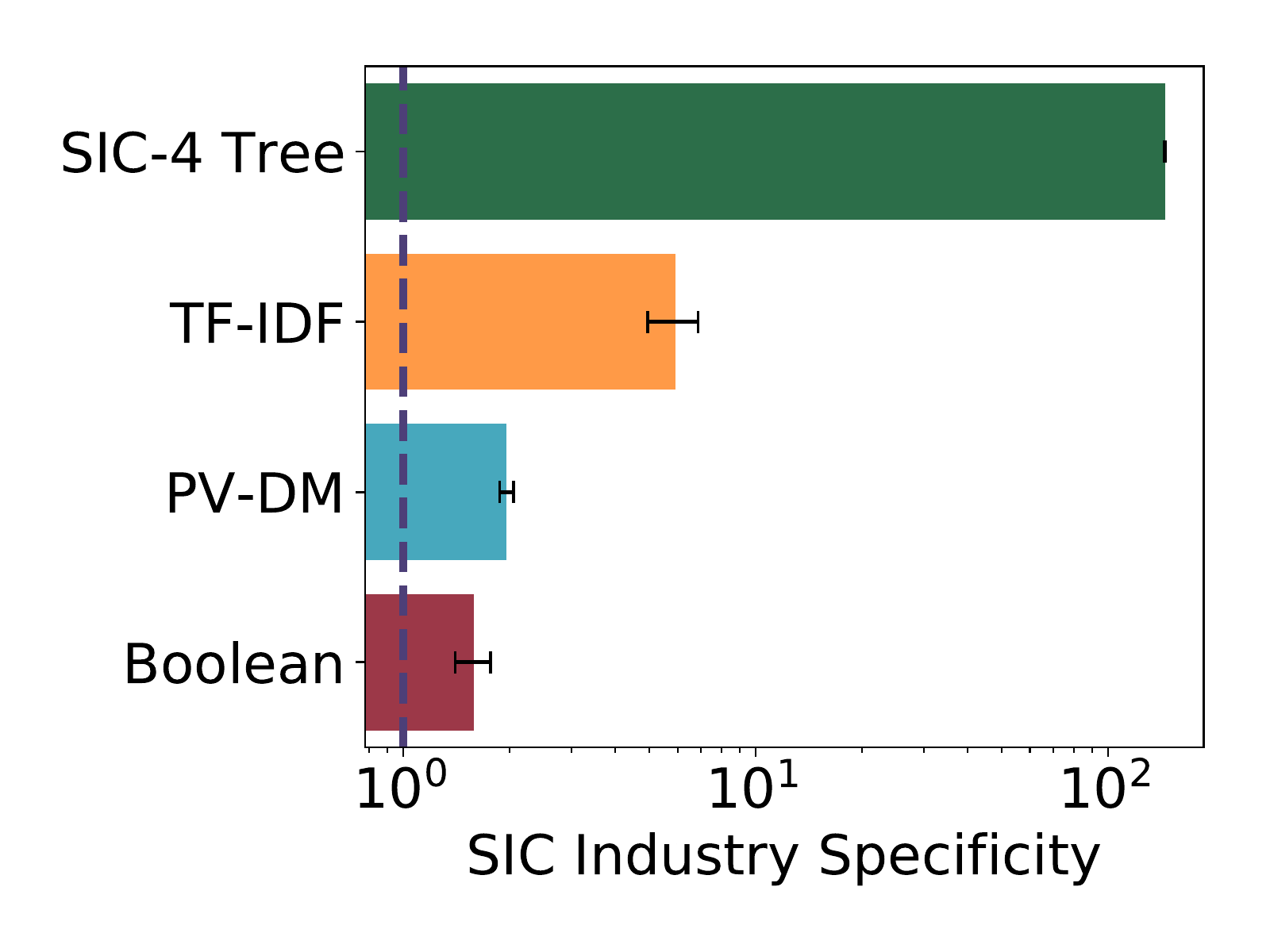}
    \caption{Industry Specificity with respect to SIC of text-based models (Boolean, TFIDF, and PVDM) and the text-free SIC control. The dotted line shows the null hypothesis of a perfectly flat similarity matrix.}
	 \label{specificity}
\end{figure}

The {\em SIC Industry Specificity} of each model (Boolean, TF-IDF, and PV-MD) is indicated in Figure \ref{specificity}. A vertical dotted line in the Figure shows the Industry Specificity of a Flat control model in which all firms are identical to one another. A model passes the SIC Industry Specificity test for plausibility if and only if it's Industry Specificity is significantly higher than the Flat control. 


Figure \ref{specificity} shows that the Specificity of the Boolean, TF-IDF, and PV-MD models all is significantly higher than the Flat control. The three text-based models all construe firms from the same SIC class as more similar on average than firms from different classes. All of the text-based embeddings have higher firm similarities on average for pairs of firms with the same SIC Code. Thus, if the SIC in fact reflects some genuine structure among all the industries, the text-based vector embeddings reflect a similar structure.

Although a plausible model must have significantly more Industry Specificity than the Flat control, higher industry specificity is not always better. A model's Industry Specificity reflects the degree to which its similarity matrix corresponds to some trusted reference classification (here, SIC Codes). But there is no guarantee that the reference classification captures all the relevant observable details about the actual similarities among firms. A high Industry Specificity could reflect a model's high correspondence to a trusted reference's crude model of firm similarity. Furthermore, a model's level of Industry Specificity is roughly correlated with the off-diagonal heat in it's firm similarity matrix, shown in Figure~\ref{heatmaps}. But there is no reason to think that better models have lower off-diagonal heat, for there certainly is some similarity of some firms that are far apart in the SIC classification tree. So, any plausible model should have an SIC Industry Specificity that is significantly higher than the Flat control, but Industry Specificity should not be viewed as a model success metric. 




\section{Product diversity trends}

We measure the diversity of the products of large US firms from the past twenty years (1997 to 2017), and look for trends over these years. We have already measured the annual diversity of products of large US firms simply as the number of four-digit SIC Codes that are instantiated each year (Figure~\ref{SIC4} left) and as the Shannon entropy of the distribution of SIC Code counts each year (Figure~\ref{SIC4} right). We now focus on diversity measurements that depend on annual individual firm-level similarity matrices, produced by embedding in a semantic vector space a description of each firm's products in a given year. We compare the trends in product diversity disclosed by embedding firms in the Boolean, TF-IDF, and PV-DM vector spaces.

\begin{figure*}[h]
	       \centering
    \includegraphics[width=0.32\linewidth]{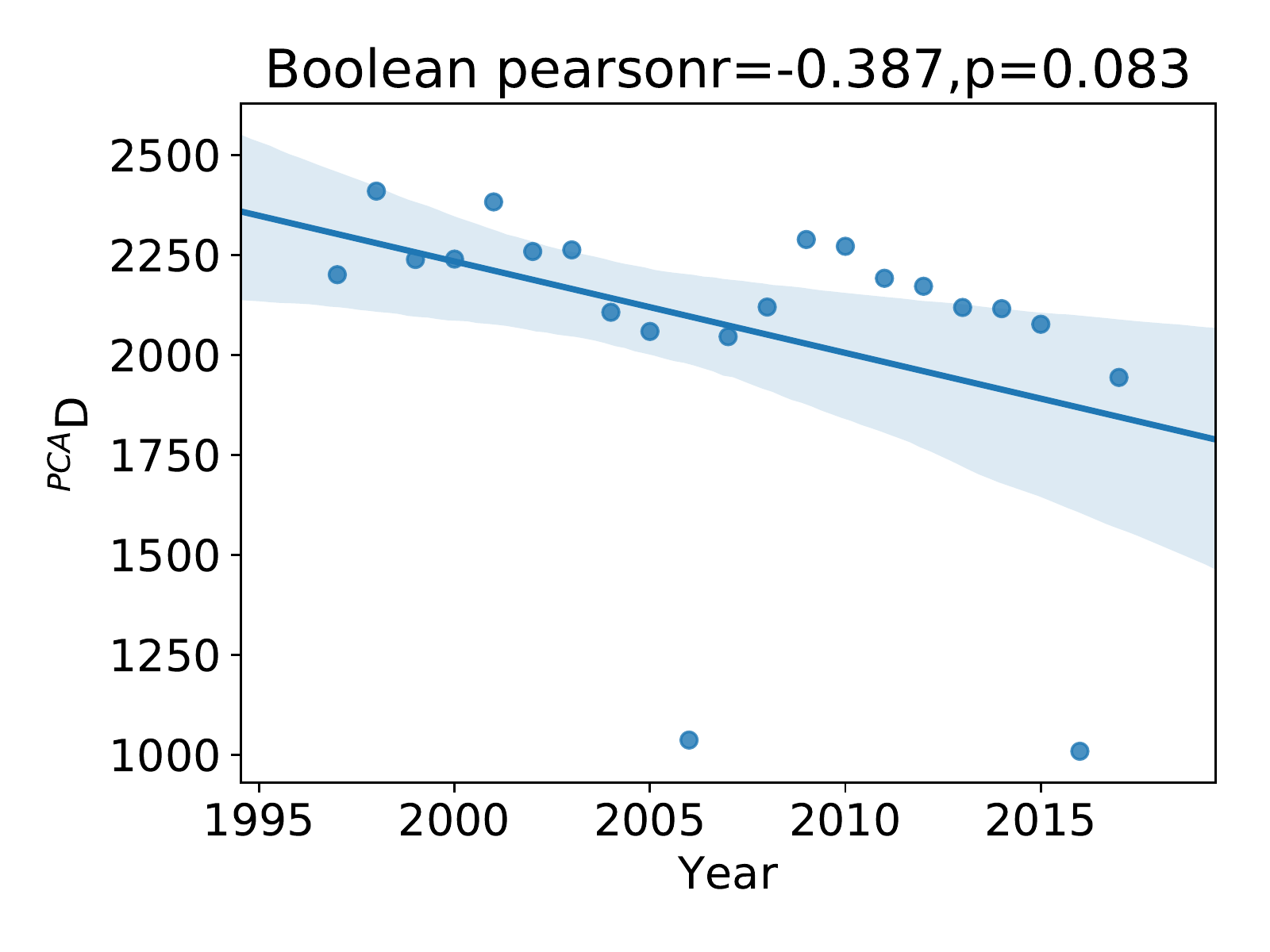}
  \includegraphics[width=0.32\linewidth]{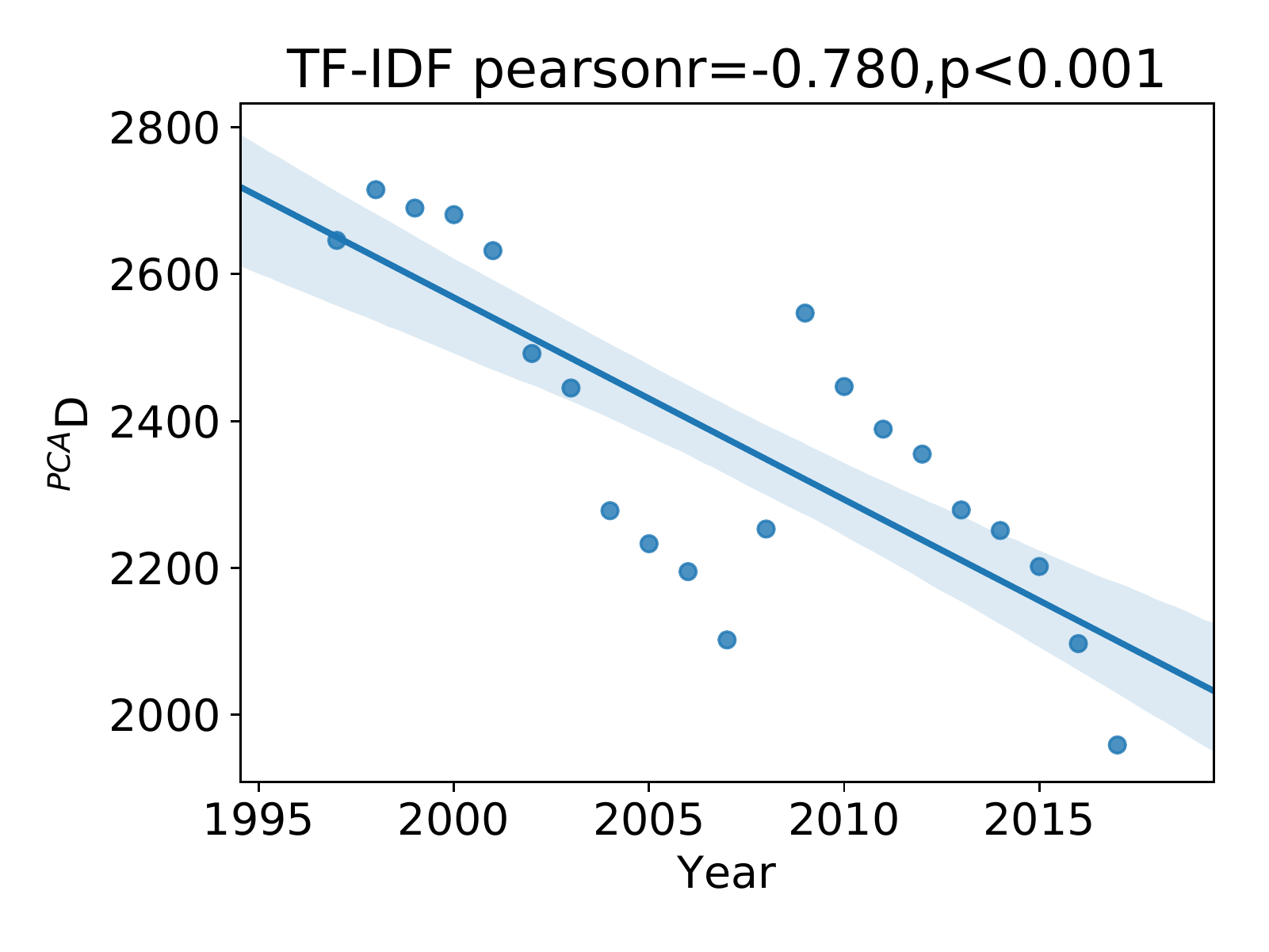}
  \includegraphics[width=0.32\linewidth]{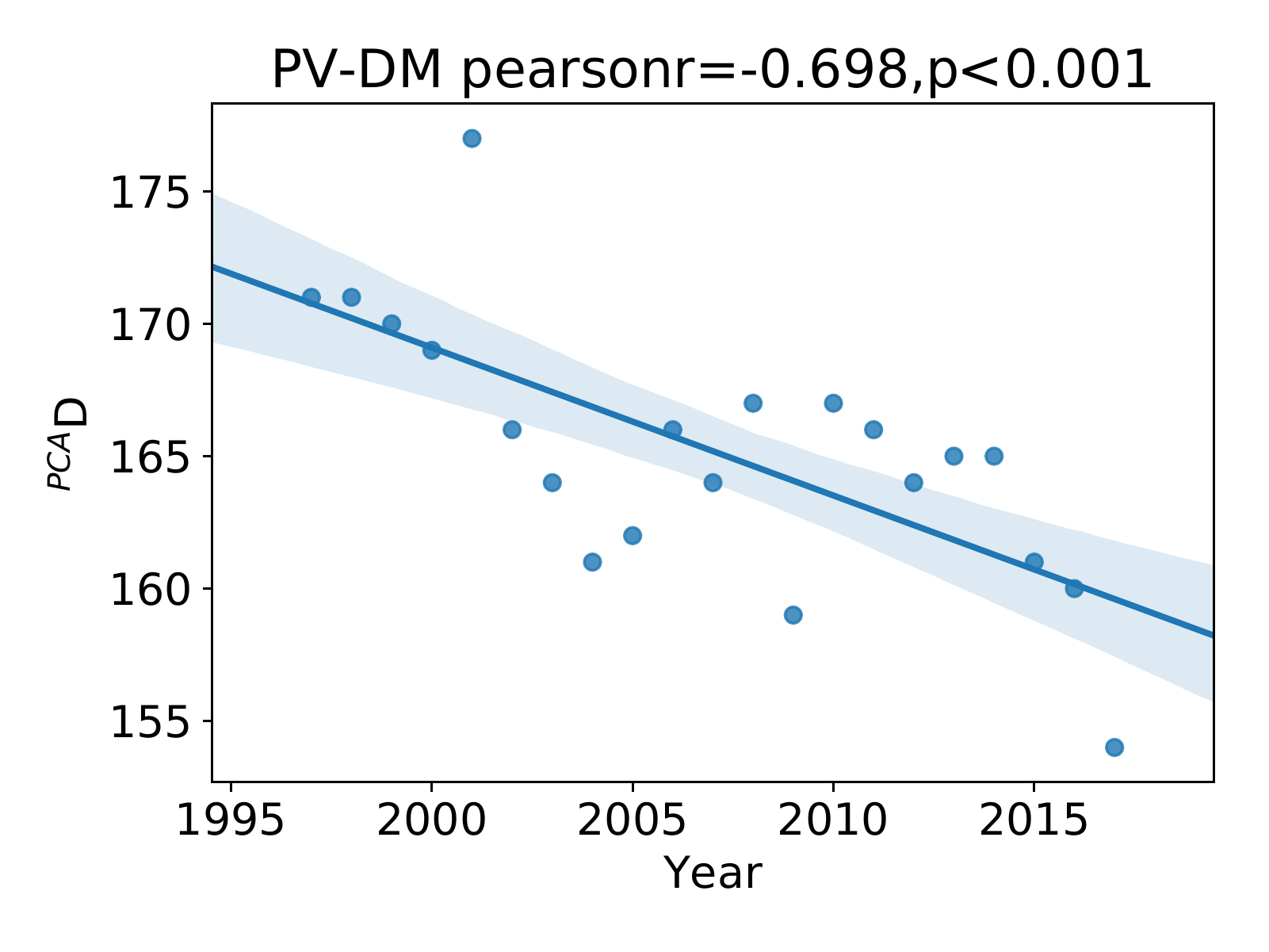}
    \caption{Annual number of dimensions required to account for 90\% of variance of document vectors when embedded in the Boolean (left), TF-IDF (middle), and PV-DM (right) models. For each model we show a scatter plot of annual diversities and a linear regression fit with its 90\% confidence interval.}
	 \label{PCA}
\end{figure*}

One measure of diversity, $^{PCA}$D, is simply the number of principle components required to account for 90\% of the variance in the spread of firms embedded each year. Figure \ref{PCA} shows $^{PCA}$D results for all three models, illustrating a consensus drop in diversity over the period of analysis. The Boolean and TF-IDF models show almost a 30\% drop in $^{PCA}$D over twenty years, while the drop shown by the PV-DM model is only 8\%. 

\begin{figure*}[h]
	       \centering
    \includegraphics[width=0.32\linewidth]{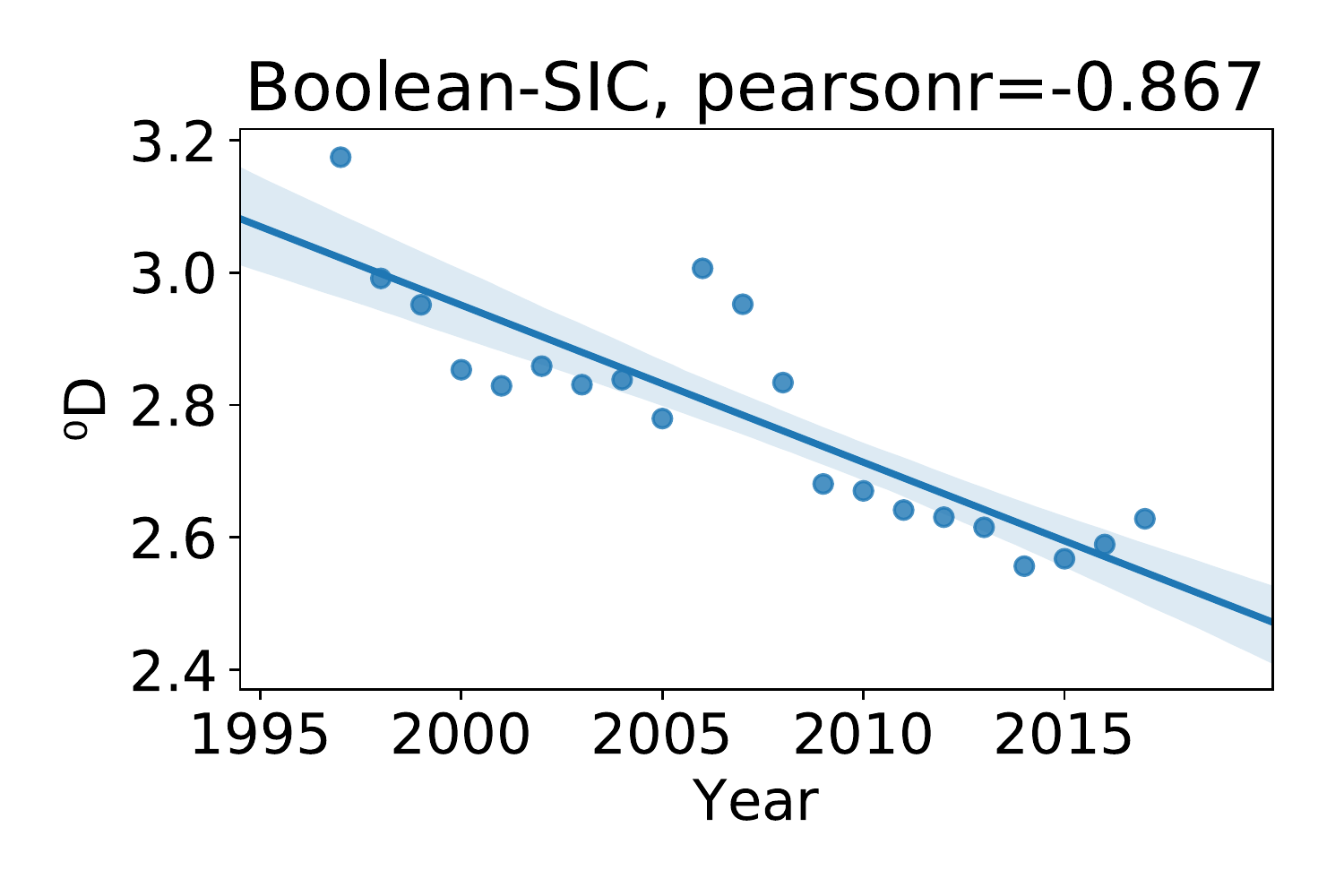}
  \includegraphics[width=0.32\linewidth]{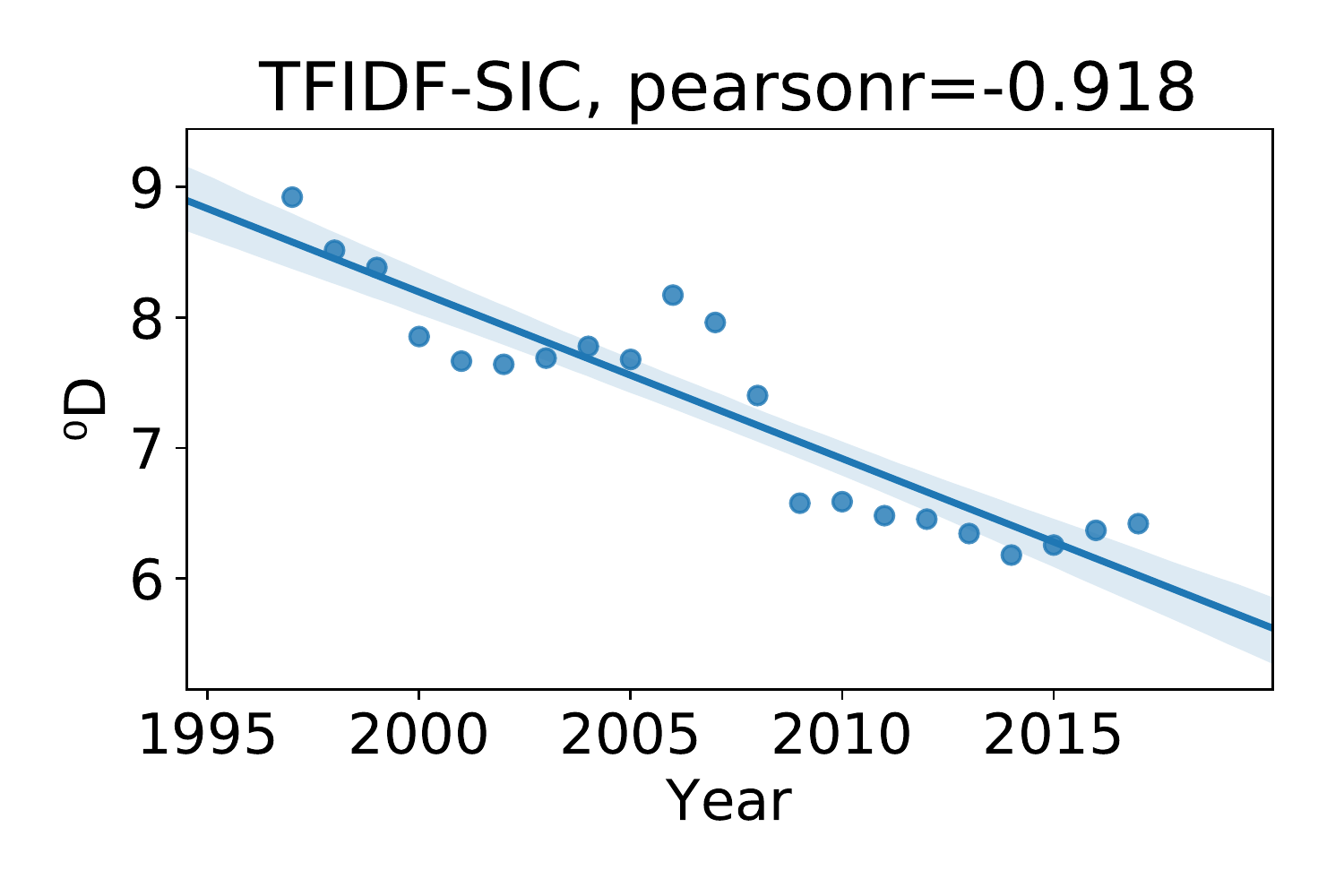}
  \includegraphics[width=0.32\linewidth]{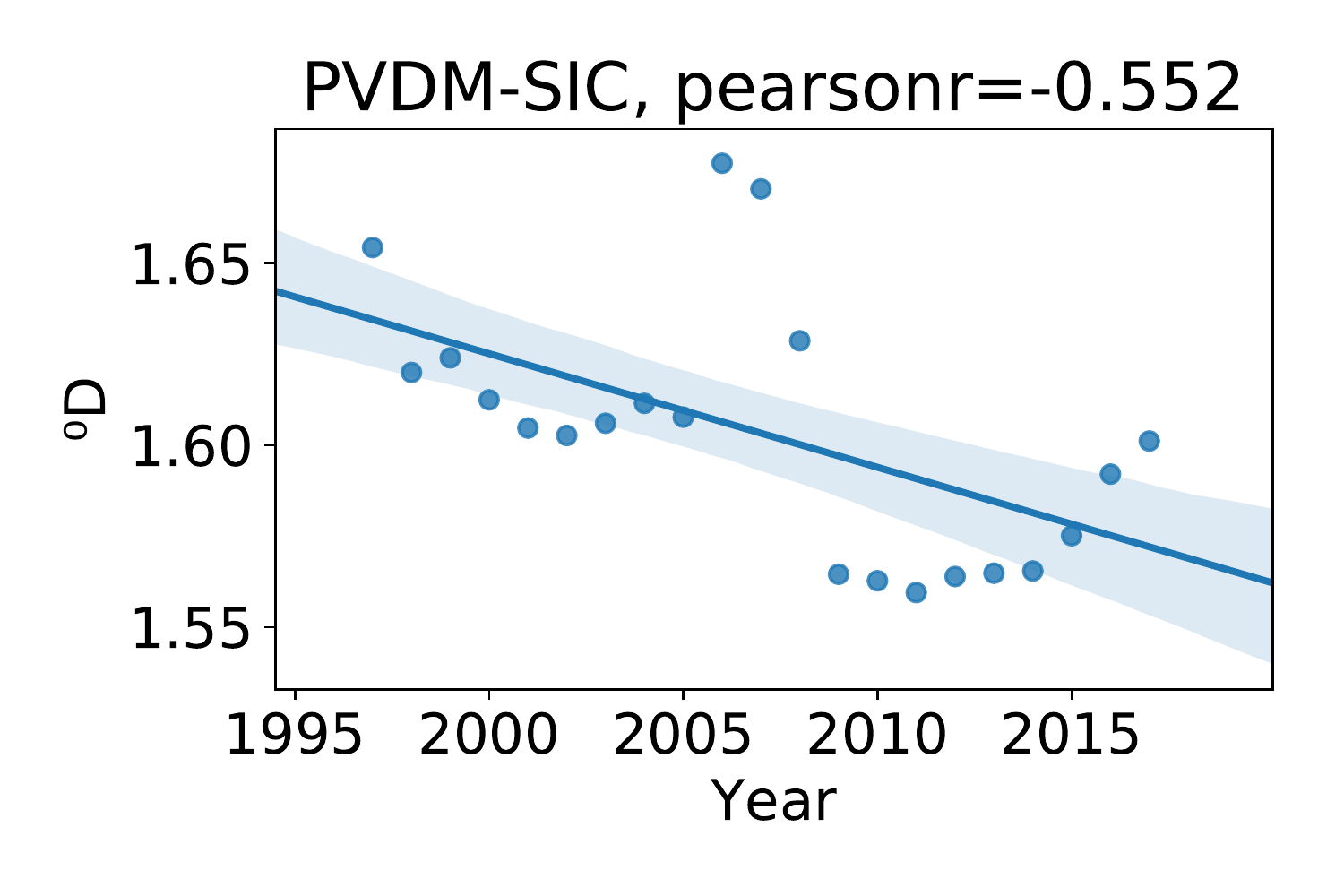}

    \caption{Annual economic diversity of order zero, $~^0D$,  according to models Boolean (left), TF-IDF (middle), and PV-DM (right). Each model includes a scatter plot of diversities and a linear regression fit along with its 90\% confidence interval.}
	 \label{diversity}
\end{figure*}

\begin{table}
\centering
\scalebox{1.0}{
\begin{tabular}{llll}
  \hline
 & $~^0D$& $~^2D$ & $~^5D$ \\ 
  \hline
  Boolean & $-0.867^{***}$ & $-0.856^{***}$ & $-0.820^{***}$\\
  TF-IDF & $-0.918^{***}$ & $-0.894^{***}$ & $-0.808^{***}$\\ 
  PV-DM & $-0.552^{***}$& $-0.543^{**}$ & $-0.531^{**}$ \\
   \hline
   \vspace{-.1in}
\end{tabular}
}
\caption{Correlation coefficients of diversity $~^qD$ with year and significance levels ($~^{**}$: p-value $\leq0.05$, $~^{***}$: p-value $\leq0.01$).}
\label{diversity_corr}
\end{table}

In order to understand the effect on diversity trends of varying the degree of sensitivity to rare species, annual $~^qD$ values are calculated for $q\in\{0,2,5\}$. These diversity values reflect not just the abundances of different SIC classes but also how similar the classes are to each other. The scatterplots of annual diversity values and linear regression fits for $q=0$ and the three models of interest are shown for on Figure~\ref{diversity} while the Pearson correlation coefficients for all the tested sensitivities are shown on Table~\ref{diversity_corr}.

Table \ref{diversity_corr} and Figure~\ref{diversity} show that for $q=0$, all three models show statistically significant decreasing trends in diversity. This means that all three models are in agreement that the richness of products is decreasing over the years. In other words, \emph{the trend of dropping product richness} in the (descriptions of) products of large US firms is a consensus conclusion of diversity measurements with $q=0$ of Boolean, TF-IDF and PV-DM models. Similarly, all three text-based models show statistically significant decreasing trends in $~^2D$. $~^2D$ starts to pay less emphasis to rare species and is equivalent to a commonly used diversity measure in ecology known as Rao's quadratic entropy \cite{sim_diversity}. Finally, further increasing $q$ to 5 continues the pattern of Boolean, TF-IDF and PV-DM showing statistically significant decreasing trends in diversity. The upshot of the diversity correlation coefficients is that all models show statistically significant patterns of dropping diversity across different sensitivity values. \added[id=AN]{As with Shannon entropy, to take away the effect of the decreasing number of SIC Codes on $~^qD$, the metric can be normalized as described in S2 Appendix. It can be seen that the normalized $~^0D$ also shows a decreasing trend in the Boolean and TF-IDF models and no significant trend with PV-DM.}

\section{Conclusions and Discussion}

This paper presents a wealth of evidence for a significant drop in the diversity of the products produced by large US firms in this century. This downward trend is evident whether diversity is measured in crude or sophisticated ways, and whether the information about the products of individual firms is coarse- or fine-grained. This trend can be seen using a Boolean word vector model, the current industry standard in product-focused firm embeddings due to Hoberg and Phillips \cite{hoberg2016text}, and the trend can be seen using more sophisticated TF-IDF and PV-DM models. The trend is even evident in a simple model based merely on a firm's four-digit SIC Code. The magnitude of the drop in diversity ranges from 6\% to 30\% depending on the method by which diversity is measured, and all the diversity measurements show some scatter year-by-year diversity, but the overall twenty-year trend of dropping product diversity is a very robust result.

Our product diversity results focus on large US firms, because our models are trained on documents that are filed only by large US firms. Since large US firms are an unrepresentative sample of all of the firms that contribute to the economy, so whether the dropping product diversity trend also holds for smaller firms and firms outside the US remains an open question. Even if we restrict our attention to large US firms, it also remains an open question how to explain the dropping diversity trend. We noted earlier an overall drop in number of firms over the same years (recall Figure~\ref{firmcounts}), and this drop in the number of firms might be thought to explain the drop in diversity of products. \added[id=TR]{ Further, it is known that since the 1990s market concentration has been occurring as fewer firms take up more market share in their industries \cite{grullon2019us,bessen2017information}.} However, we still observed the diversity drop when we measured diversity using normalized abundance vectors, so the drop in the number of firms is unlikely by itself to explain the observed trend in dropping product diversity. 

A second, quite different hypothesis is document homogenization, which proposes that the decreasing diversity of the descriptions of firms' products is due merely to an increasing professionalization and standardization of the text in 10-K documents. This hypothesis suggests that models trained on textual documents provide evidence for a drop in diversity of the {\em descriptions} of products of large US firms, but {\em not} evidence for a decrease in diversity of the products themselves. The document homogenization hypothesis does not explain why the model of firm similarity based solely on a firm's SIC Code shows a similar drop in product diversity (recall Figure~\ref{SIC4}); nor does it explain the roughly 50\% increase in average number of word tokens and word types in each document in the training corpus (recall Figure~\ref{tokensandtypes}). So the dropping diversity seen using text-based models is unlikely to be due specifically to document homogenization. 

A number of further hypotheses could explain the dropping product diversity trend. One is the hypothesis that products have shrunk in diversity because consumer demand for products has narrowed. Another hypothesis is that the growing diffusion of information technology into more and more products is making products overall more alike. A third hypothesis is that the drop in product diversity is due to the rise of outsourcing by large US firms, and a consequent \emph{rise} in the diversity of products produced {\em outside} the United States. A fourth hypothesis would connect the drop in diversity of the products of {\em large} firms with a rise in the diversity of products produced by {\em small} firms. We have no specific evidence for or against any of these hypotheses, but all of them have empirically testable consequences. However, gathering accurate and complete data about the products of firms of most sizes in most countries remains a huge hurdle.

One final hypothesis worth considering is that the trend of falling product diversity is explained by an increasing diversity of products {\em within} large US firms. On this hypothesis, the total diversity of products in the marketplace may be stable or growing, because individual large US firms on average have been producing an increasingly diverse array of products. The diversity of products produced by some individual firms has been studied, and some have grown more diverse over time. When we measure the diversity of the products produced by large US firms, the products of each firm is embedded as a point in a high-dimensional product space, and we measure the diversity of those points in product space. So, those measurements reflect the diversity {\em between} the products produced by different firms. Since all the products of an individual firm in a given year are embedded into a point in product space, the diversity measurements are blind to the diversity of products {\em within} each firm in that year. (Of course, the diversity measures do reflect changes over the years in the diversity of the product offerings of each individual firm. Recall the clusters of points with the same color in Figure~\ref{tsne}.) Nevertheless, Figure~\ref{tokensandtypes} shows that shows that there is a significant rise in the number of word tokens and word types in 10-K {\em descriptions} of the products of large US firms. This does provide some corroboration for the hypothesis that the products of individual large US firms has grown more diverse during the past twenty years.

\section*{Acknowledgments}
We thank Francois Lafond, Kieran Marray, Norman Packard, Doyne Farmer and Sergei Maslov for suggestions on the manuscript, and we thank Amir Amel-Zadeh for help with document pre-processing.
This project was accomplished with the support of project "Predicting Technological Progress by Embedding Input-Output Networks in Semantic Technological Spaces" (PR0040), in response to IARPA-BAA-17-01.




\nolinenumbers

%
%
%
\bibliographystyle{plos2015}
\bibliography{plos_latex_template}

\end{document}


\section*{Supporting information}




\paragraph*{S1 Appendix.}
\label{S1_Appendix}
{\bf Definition of Industry Specificity.}

Let $F = \{ f_1 , f_2, ..., f_n \}$ be the set of firms (semantic vectors) that exist in some particular year, and let each industry in a reference classification, $C$, be a set of firms. So, a reference classification is a set of industries (i.e., classes of firms):
$C = \{ c_1, c_2, ..., c_m \} = \{ \{ f_i, ..., f_l \} , \{ f_j, ..., f_m \}, ..., \{ f_k, ..., f_n \} \}$. Here, we let the set of firms in each 4-digit SIC Code to be our reference classification. In this case, $C$ partitions the set, $F$, of all firms (that exist in a given year), so that every firm is in exactly one class (SIC Code). Then, we can can write the {\em mean between-class similarity} of a firm similarity matrix, $M$, for a given year $Y$ as the mean similarity between pairs
of firms which do not belong to the same industry, averaged over all industries:
\begin{equation}
\sum_{i =1}^{| C |}
\left (  { { \sum_{f \in c_i , f^{\prime} \in c_j, i \neq j} {\rm sim}_M (f, f^{\prime }) }  \over { | c_i | \times ( | c_i | - 1 ) } } \right ) {1 \over  | C |}
\end{equation}
To improve readability we typically leave the year $Y$ implicit.


We can also write the mean {\em mean within-class similarity} of a firm similarity matrix $M$ in year $Y$ as the mean similarity of all pairs of firms within the same industry, averaged over all industries:
\begin{equation}
\sum_{i =1}^{| C |}
\left (  { { \sum_{f \in c_i , f^{\prime} \in c_i , f \neq f^{\prime }} {\rm sim}_M (f, f^{\prime }) }  \over { | c_i | \times ( | c_i | - 1 ) } } \right ) {1 \over  | C |}
\end{equation}
\noindent

Finally, we define the {\em Industry Specificity (relative to the SIC)} of a firm similarity matrix $M$ as its mean within-class similarity (according to $M$) divided by its mean between-class similarity; that is, the SIC Industry Specificity of $M$ is:
\begin{equation}
{
{\rm mean\ within\ SIC\ Code \ similarity\ of}\ M
\over
{\rm mean\ between\ SIC\ Code\ similarity\ of}\ M 
}
\end{equation}
The SIC industry specificity of model $M$ is the mean similarity (according to $M$) of firms within each industry (according to the SIC) expressed in units of the mean global firm similarity.

\paragraph*{S2 Appendix.} 

\label{S2_Appendix}
{\bf Adjusting for Maximum Diversity.}
It might be useful to remove the change in the number of species/industries of the ecosystem for the diversity measures that use similarity matrices.

Take the definition for diversity given by Equation 3 in the main text. Maximum diversity occurs when the similarity matrix is an identity matrix, $Z = I$ and the abundance vector $~^{balanced}a$ is perfectly balanced such that $a_{y,1}=...=a_{y,s}=\frac{1}{s}$. Then,
\begin{align}
    ~^qD(~^{balanced}a,I) &= (\sum_i^s a_{i} (\sum_j^s I_{ij}a_{j})^{q-1} )^\frac{1}{1-q}\\
    &= (\sum_i^s a_{i} (a_i)^{q-1} )^\frac{1}{1-q}\\
    &= (\sum_i^s (\frac{1}{s})^{q} )^\frac{1}{1-q}\\
    &= s.
\end{align}
Therefore, the adjusted diversity of order q is given by
\begin{align}
    ~^qD_{adj}(a,Z) &= \frac{~^qD(a,Z)}{s}.
\end{align}

\begin{figure*}[!h]
	       \centering
    \includegraphics[width=0.32\linewidth]{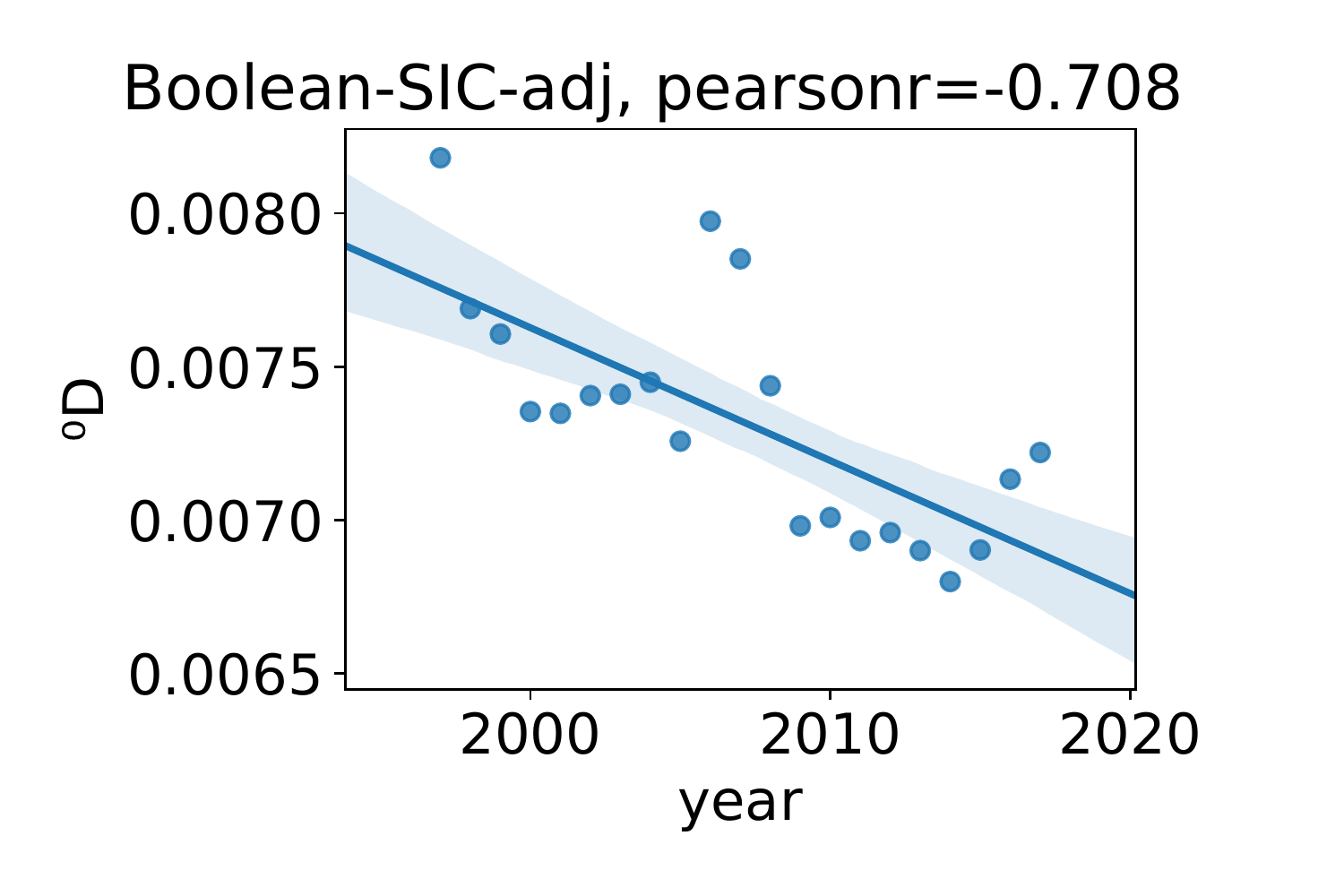}
  \includegraphics[width=0.32\linewidth]{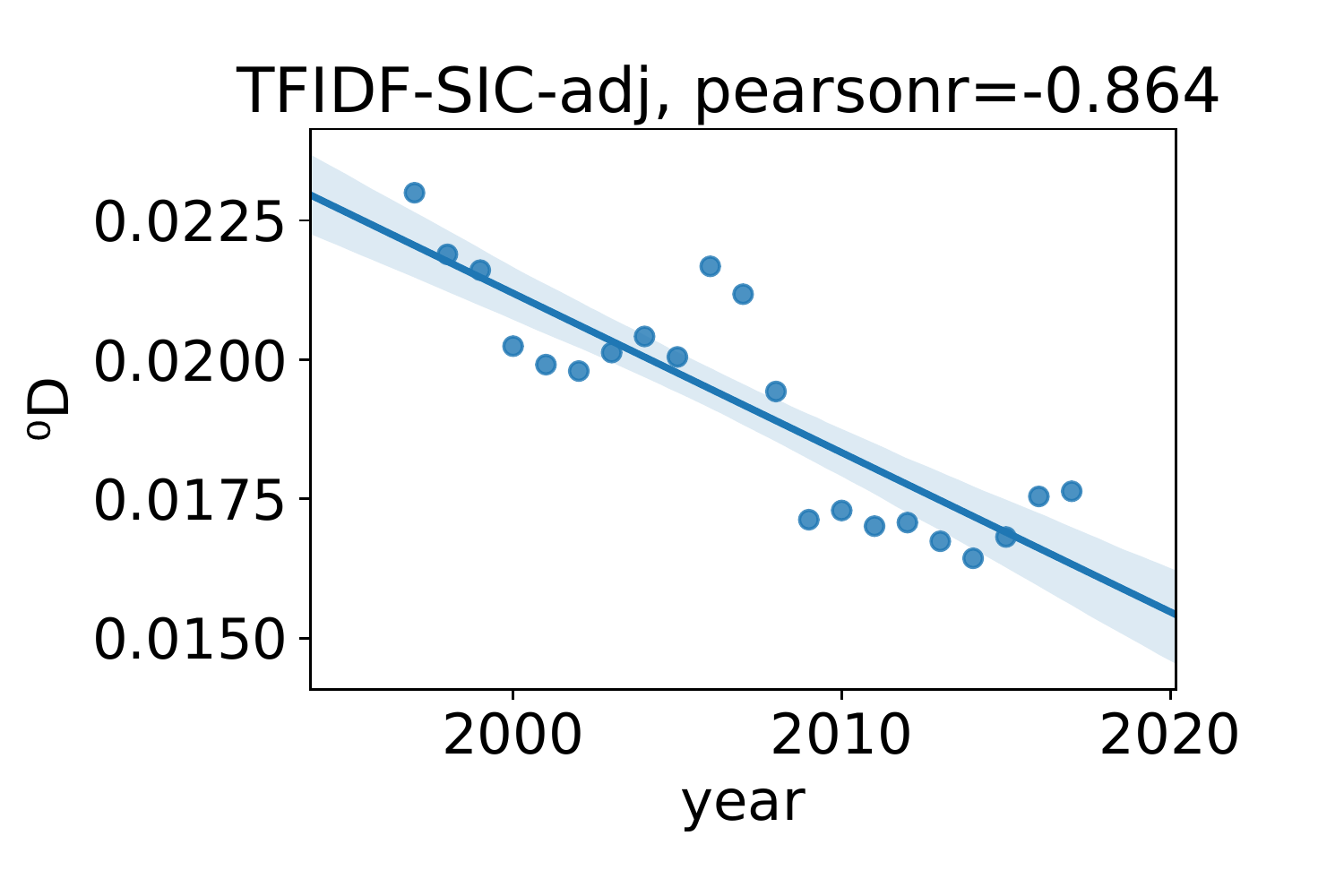}
  \includegraphics[width=0.32\linewidth]{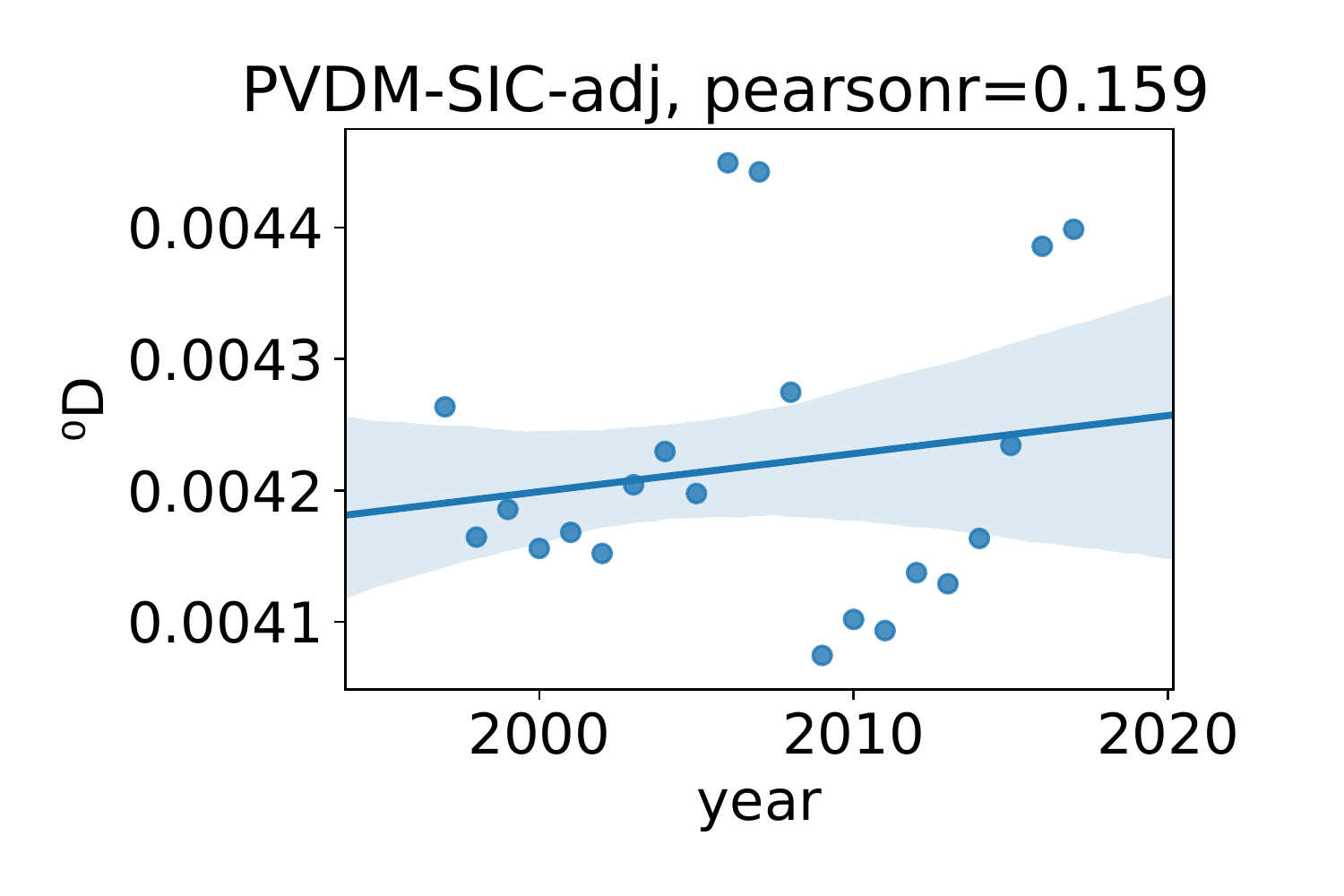}
  
    \caption{Annual economic diversity of order zero, $~^0D$, adjusted for varying  number of the 4-digit SIC Codes that are instantiated, according to models: Boolean (left), TF-IDF (middle), and PV-DM (right). Each model includes a scatter plot of diversities and a linear regression fit along with its 90\% confidence interval.}
	 \label{diversity_norm}
\end{figure*}